\documentclass[12pt, draftclsnofoot, onecolumn]{IEEEtran}
\IEEEoverridecommandlockouts
\usepackage{verbatim}
\usepackage[utf8x]{inputenc}
\usepackage{epstopdf}
\usepackage[pdftex]{graphicx}
\usepackage{tabularx}
\usepackage{booktabs}
\usepackage[table]{xcolor}
\usepackage{booktabs}
\usepackage{ctable}
\usepackage[cmex10]{amsmath}
\usepackage{multicol}
\usepackage{algorithm}
\usepackage{algpseudocode}
\usepackage{amssymb}
\usepackage{cite}
\usepackage{array}
\usepackage{caption}
\usepackage{subcaption}
\def\BibTeX{{\rm B\kern-.05em{\sc i\kern-.025em b}\kern-.08em
		T\kern-.1667em\lower.7ex\hbox{E}\kern-.125emX}}

\usepackage[letterpaper, top=.75in, bottom=.95in,left=0.625in, right=0.625in]{geometry}

\begin{document}
\bstctlcite{IEEEexample:BSTcontrol}
\title{Distributed Wideband Sensing-based Architecture for Unlicensed Massive IoT Communications} 

\author{Ghaith Hattab,~\IEEEmembership{Student Member,~IEEE,} Danijela Cabric,~\IEEEmembership{Senior Member,~IEEE}
	\thanks{This paper was presented in part at the IEEE Global communications Conference (Globecom), Abu Dhabi, the UAE, December 2018 \cite{HattabCabric2018a}.
	G. Hattab and D. Cabric are with the Department of Electrical and Computer Engineering, University of California, Los Angeles, CA 90095-1594 USA (email: ghattab@ucla.edu, danijela@ee.ucla.edu).}
}

\maketitle

\begin{abstract}
Providing Internet connectivity to a massive number of Internet-of-things (IoT) objects over the unlicensed spectrum requires: (i) identifying a very large number of narrowband channels in a wideband spectrum and (ii) aggressively reusing the available channels over space to accommodate the high density of IoT devices.  To this end, we propose a sensing-based architecture that identifies spectral and spatial resources at a fine resolution. In particular, we first propose a sensing assignment scheduler, where each base station (BS) is assigned a subset of the spectrum to sense at a high resolution. We then propose a distributed sensing algorithm, where BSs locally process and share their sensing reports, so that each BS obtains occupancy information of the wideband spectrum at its location. Once the spatio-spectral resource blocks are identified, we further propose a distributed resource allocation algorithm that maintains high spatial reuse of spectral opportunities while limiting the intra-network and inter-network interference. Numerical simulations are presented to validate the effectiveness of the proposed distributed algorithms, comparing them to centralized and non-cooperative schemes. It is shown that our architecture identifies more spatio-spectral resources, with lower misdetection of incumbents. As a result, more IoT devices are connected with limited interference into incumbents. 
\end{abstract}


\section{Introduction}
Providing wireless Internet connectivity to a massive number of sensors and machines, collectively known as the \emph{Internet-of-things} (IoT) objects, brings a myriad of applications across many vertical sectors such as smart cities \cite{Zanella2014}, public safety \cite{Cisco2014}, and agriculture \cite{Microsoft2016}. The transformative societal impact and the economic benefits massive IoT brings have led to new use-cases in fifth-generation new radio (5G-NR) \cite{3GPP2015a,ITU2015} and new  emerging technologies such as low-power wide-area (LPWA) networks \cite{Raza2017,CentenaroZorzi2016}. 

The high density of devices in massive IoT applications has made \emph{spectrum sharing} a key challenge to fully realize large-scale deployment of these low-cost low-rate devices. For licensed-based access, cellular networks have introduced new user categories such as LTE-M and narrowband IoT (NB-IoT) \cite{Rico-AlvarinoYavuz2016,Nokia2016}, which aim to reduce device complexity and connect more devices over narrowband channels. In addition, cellular networks rely on congestion control methods to ensure fair coexistence between cellular users and IoT devices \cite{Jiang2018a,DawyYaacoub2017}. Such access control requires owning a licensed spectrum, which is expensive. For this reason, a growing number of IoT-based networks have centered around the use of the unlicensed spectrum \cite{Raza2017,MulteFire2017}, which is the focus of our work.

\subsection{Related Work}
Due to the presence of various wireless technologies in the unlicensed bands, spectrum sharing is more challenging, particularly, when the low-cost IoT device is required to coexist with more powerful end-devices, e.g., WiFi devices. Thus, the vast majority of LPWA networks rely on narrowband signals to improve robustness to interference and to connect a large number of devices \cite{CentenaroZorzi2016}. For example, in LoRa, spread-spectrum signals have bandwidth of 125KHz-500KHz \cite{SorninHersent2016}, whereas Sigfox uses ultra-narrowband signals with bandwidth 100-600Hz \cite{Sigfox2017}. While LPWA networks can achieve wide coverage and connect a very large number of devices \cite{Bor2016,HattabCabric2018b}, they have several limitations. First, they are constrained to very low-rate applications, with data rates below 50Kbps \cite{Raza2017}. Second, they primarily operate over the sub-1GHz band due to its favorable propagation conditions, yet it has limited bandwidth, e.g., 26MHz at 915MHz in the US. Third, LPWA networks rely on ALOHA-like access without any listen-before-talk (LBT) or channel clear assessment (CCA) mechanisms. Thus, they cannot be deployed at several unlicensed bands because of the regulatory requirements. For example, Japan and Europe mandate LBT for access at 5GHz, whereas dynamic frequency selection is required in some bands by the Federal Communications Commission (FCC) in the US \cite{Kwon2017}. In this work, we explore the use of sensing-based access for unlicensed massive IoT as wider bands are available at 2.4GHz and 5GHz.

One prime example of a network that relies on LBT for IoT access is the MulteFire network \cite{MulteFire2017}, which is a 3GPP-compliant standard for cellular-like access in the unlicensed spectrum \cite{Labib2017}.\footnote{MulteFire differs from licensed-assisted access (LAA)  as it does not require any anchor in the licensed band \cite{Kwon2017}.} It combines the simplicity of WiFi deployment and the reliability of cellular networks, and it is envisioned to enable a stand-alone unlicensed 5G-NR \cite{3GPP2018}. The authors in \cite{Li2016,Cano2017} theoretically analyze the coexistence of WiFi and LTE with LBT, showing that both systems benefit from such coexistence. In \cite{Rosa2018}, system-level simulations show that MulteFire outperforms WiFi in coverage and capacity. Alternative to using LTE-like IoT networks, the authors in \cite{Rajandekar2017} propose to modify the MAC protocol for WiFi itself, where IoT devices are notified by the WiFi access point (AP) about available \emph{WiFi white spaces} when the latter no longer has packets from/for WiFi devices. The presence of multiple APs is not considered, which could affect the scalability of this protocol. Finally, the aforementioned works, and the MulteFire network, merely support private IoT networks, e.g., for enterprise in-building cases, and thus they do not necessarily scale to support massive IoT unless each base station (BS) or AP is equipped with a wideband spectrum scanner. In this work, we develop a sensing assignment scheduler so that each BS learns the occupancy of a wideband spectrum using narrowband scanners.

Other works have proposed the use of compressive sensing techniques to enable low-cost wideband scanners for IoT networks \cite{Ejaz2018,Zhang2018}. Nevertheless, cooperation among sensing scanners destroys spatial information about incumbents' energy footprints, i.e., all cooperating scanners arrive at the same sensing decision, and so available channels cannot be reused across space. In this work, we propose a truly distributed sensing algorithm, where cooperating scanners can arrive at different decisions, capturing the spatial variations of incumbents' footprints. Finally, the authors in \cite{Khan2017} propose the use of radio environment maps (REMs) so that different IoT networks share the spectrum with rotating radars. The framework, however, requires radar operators to share radar locations and operations with a REM repository. In our work, we show that the output of the distributed sensing algorithm resembles a REM, which is obtained without any cooperation with incumbents.

\subsection{Contributions}
In this paper, we propose a sensing-based architecture for unlicensed massive IoT, where BSs, e.g., small cells, are equipped with spectrum scanners to: (i) identify a large number of narrowband channels in a wideband spectrum, as many massive IoT applications have low-rate requirements, and to (ii) aggressively reuse the unlicensed channels over space to accommodate a high density of IoT devices. The former objective requires sensing at a fine \emph{spectral} resolution, and the latter requires capturing the incumbents' footprints at a fine \emph{spatial} resolution. 

Our contributions in this work are twofold. First, we aim to limit the sensing burden at each BS, as sensing a wideband spectrum at a fine spectral resolution requires complex receivers. To this end, we formulate an integer program in Section \ref{sec:scheduler}, where we optimize sensing assignments across BSs, so each one only senses a subset of the spectrum. The problem is combinatorial with high complexity in dense networks. Thus, we further develop a heuristic low-complexity assignment scheduler, and compare its performance to the integer program and its linear relaxation. We note that sensing assignments have been studied before in a different context in \cite{Mochaourab2015,Lai2015,Zhang2015}. For instance, in \cite{Mochaourab2015,Lai2015}, the assignment is done such that each channel is sensed by one device, whereas in this work we require each channel to be sensed by multiple BSs for reliable cooperative sensing. In \cite{Zhang2015}, the assignment aims to maximize the rate of secondary users, and thus it requires these users to sense all channels before making the assignment. In this paper, the assignment is done prior to sensing, where we aim to ensure that each BS, sensing a specific subset of channels, is surrounded by BSs sensing other subsets. 

The second contribution is the development of distributed sensing and resource allocation algorithms. In particular, we propose, in  Section \ref{sec:sensing}, a distributed sensing algorithm, where each BS senses its assigned subset of channels, shares and collects measurements from nearby BSs, and processes the collected data to infer the spectrum occupancy across all channels. Different from the distributed sensing algorithm in \cite{SobronVelez2015}, each BS may arrive at a different decision as the occupancy of a channel vary over space, achieving high spatial resolution. In addition, we use the combine-then-adapt diffusion algorithm \cite{ChenSayed2015} and propose a novel update of the algorithm's weights to quickly diffuse information about the wideband spectrum at each BS. The proposed sensing algorithm is shown to improve the deflection coefficient \cite{Unnikrishnan2008} compared to the non-cooperative energy detection. In fact, the algorithm weights resembles a REM of incumbents' footprints. The output of the sensing algorithm determines which channels are available at each BS, henceforth denoted as \emph{spatio-spectral blocks}. This output is then used to maximize a network utility function by optimizing which spatio-spectral resources are allocated across BSs, so that high spatial reuse is achieved while still limiting the intra- and inter-network interference. The network optimization, presented in Section \ref{sec:access}, is centralized, and thus we develop a fast distributed resource allocation algorithm, where BSs locally coordinate with their neighbors the allocation of the spatio-spectral blocks. 

We validate the effectiveness of the proposed architecture via Monte Carlo simulations in Section \ref{sec:simulations}. Specifically, we compare the distributed sensing algorithm with centralized and non-cooperative sensing. We show that the proposed approach finds more spatio-spectral blocks, with a lower misdetection in comparison with existing methods. We show that the distributed allocation algorithm performs relatively well compared with its centralized implementation, and it provides better coexistence compared to non-cooperative allocation in terms of the number of devices served and the interference-to-noise ratio (INR) at incumbents. In addition, we simulate a large-scale case study, where a massive IoT network shares a wideband spectrum with hundreds of outdoor WiFi APs. It is shown that the proposed system helps serve significantly more IoT devices compared to the non-cooperative scheme. Finally, a discussion of the computational complexity and the implementation of the proposed algorithms is presented throughout the paper. 

\emph{Notation:} In this paper, we use boldface small and capital letters to denote column vectors and matrices, respectively, and use normal font for scalars. We use $\mathbf{1}_{K}\in\mathbb{R}^K$ to denote the vector of all ones, $\mathbf{I}$ to denote the identity matrix, and $\operatorname{I}_{(x)}$ to denote the indicator function for the argument $x$. Finally, we use $\mathbb{E}[\cdot]$ and $\mathbb{V}(\cdot)$ for the expectation and variance operations, respectively. 

%

\section{System Model and Proposed Architecture}\label{sec:model}

\subsection{Network topology}

\subsubsection{IoT network}
We consider an IoT network that consists of a very large number of IoT devices and a dense deployment of $K$ BSs that are connected to the same core network. We denote the set of BSs by $\mathcal{K}=\{1,2,\cdots,K\}$. When cooperation is enabled across BSs, then each $k$-th BS can exchange data with neighboring BSs, defined by the set $\mathcal{N}_k$. We assume that any BS within distance $R$ from the $k$-th BS belongs to $\mathcal{N}_k$. We further assume  $k\in\mathcal{N}_k$. 

IoT devices are randomly deployed over space, and each device connects to the nearest BS. In this paper, we assume a fully-loaded network, i.e., each BS has IoT devices connected to. In case a channel can be accessed by a BS, the BS will schedule its associated IoT devices in a round-robin manner, although the framework can consider other scheduling and admission control algorithms.

\subsubsection{Spectrum sharing with incumbent networks}
The IoT network shares an unlicensed wideband spectrum, of bandwidth $B$, with different incumbent networks, e.g., WiFi, LoRa, etc. Since the majority of massive IoT applications have low-rate requirements, we further assume that the spectrum can be divided into narrowband channels, each of bandwidth $b\ll B$. The set of channels is denoted by $\mathcal{M}=\{1,2,\cdots,M\}$, where $M=\lfloor\frac{B}{b}\rfloor\gg1$. Incumbent devices can occupy any part of this spectrum, and each one can use many contiguous narrowband channels.

\subsubsection{Example}
An example of such coexistence of two networks over the unlicensed spectrum  is a cellular network sharing the spectrum with WiFi over the unlicensed spectrum at 5GHz, where $B\approx500$MHz. In this case, the cellular network deploys $K$ small-cell BSs with LAA or with MulteFire specifications.  The cellular network provides IoT-based services using LTE-M or NB-IoT operations \cite{Rico-AlvarinoYavuz2016}. The former requires channels of bandwidth $b=1.4$MHz, and thus $M\approx 350$, and the latter requires $b=180$KHz, i.e., $M\approx 2775$. WiFi devices can use different channels with bandwidth ranging from 20MHz to 160MHz. We note further that the FCC has opened an inquiry on the use of the 5.9GHz-7.1GHz spectrum for 5G-NR \cite{FCC2017b}, i.e., in this case $B=1.2$GHz  and $M\approx 6666$ for NB-IoT operation. 

\subsection{Energy-based sensing model}
We assume that BSs have spectrum sensors, where energy-based sensing is used at each BS. We note that unlicensed cellular standards, e.g., LAA or MulteFire, all require BSs to perform LBT prior to channel access \cite{Labib2017}. In particular, if the energy of a channel exceeds a predetermined threshold $\tau$ (e.g., $\tau=-72$dBm \cite{MulteFire2017} or $\tau=-62$dBm \cite{MukherjeeLarsson2016}), then the BS initiates a random back-off procedure or defers for an extended CCA period \cite{Kwon2017}. 
Let $Y_{k,m}$ be the energy of the $m$-th channel measured by the $k$-th BS. Then, we use the following received energy model \cite{Laghate2017}
\begin{equation}
\label{eq:MultibandDetectionProblem}
Y_{k,m}= V_{k,m}	+ \sum_{j} S_{j,k,m},
\end{equation}
where $V_{k,m}$ is the noise power and $S_{j,k,m}$ is the received signal power from an $j$-th incumbent transmitter over the $m$-th channel. Without loss of generality, we assume noise to be Gaussian, and independent at each channel and BS. For instance, the average noise power over a 20MHz channel is $\mathbb{E}[V_{k,m}]\approx -101$dBm. For the power received from an incumbent, we primarily use the 3GPP outdoor channel model that includes Rayleigh small-scale fading, shadowing, and different path losses depending on whether the incumbent-BS link is line-of-sight (LOS) or non-LOS  \cite[Sec. 7.4]{3GPP2017d}. Such model captures the variations of the received signal over space, i.e., two nearby BSs should receive correlated energy measurements, when fading is averaged out over multiple observations.

\subsection{Proposed sensing-based architecture} 
The objective of the proposed architecture is to help a network of dense BSs identify many narrowband channels in a wideband spectrum, and reuse available channels over space. To this end, to achieve a high spatio-spectral resolution map of available resources, the architecture consists of three inter-connected components, which are described next. 
\subsubsection{A sensing assignment scheduler}
Sensing a wideband spectrum, e.g., $B=500$MHz, at a fine spectral resolution, e.g., $b=180$KHz, requires high-complexity receivers, which incurs higher costs in dense network deployments. Therefore, we propose a sensing assignment scheduler, where the $k$-th BS senses $p_k\ll M$ channels to reduce the sensing burden. For instance, for NB-IoT operation, if each BS senses 20MHz out of the 500MHz spectrum, we have $p_k=111$. To improve the reliability of sensing decisions, it is desired to have each $m$-th channel sensed by $q_m$ BSs. To summarize, the assignment scheduler aims to solve the following problem: How to assign sensing tasks to a dense network of BSs so that each one senses only part of the spectrum, and each channel is sensed by multiple BSs? 
\subsubsection{A distributed wideband sensing algorithm}
Once the sensing assignments are completed, each BS locally senses the assigned channels, measuring their received energies. Given that each BS has sensing reports only for their assigned channels, a key problem is how should BSs cooperate and process sensing data across the network so that each one learns the occupancy of the wideband spectrum. To limit the overhead with the core network, the proposed sensing algorithm is implemented in a distributed manner, where data exchange is done only among neighboring BSs. Such algorithm requires an initial phase, where the $k$-th BS measures the power of a reference signal broadcasted by the $j$-th neighboring BS, which is denoted by $\hat P_{k,j}\forall j\in\mathcal{N}_k$. Such measurement can be used to assess the quality of the sensing reports received from neighboring BSs.
\subsubsection{A distributed resource allocation algorithm}
\begin{figure}[t!]
	\center
	\includegraphics[width=3.5in]{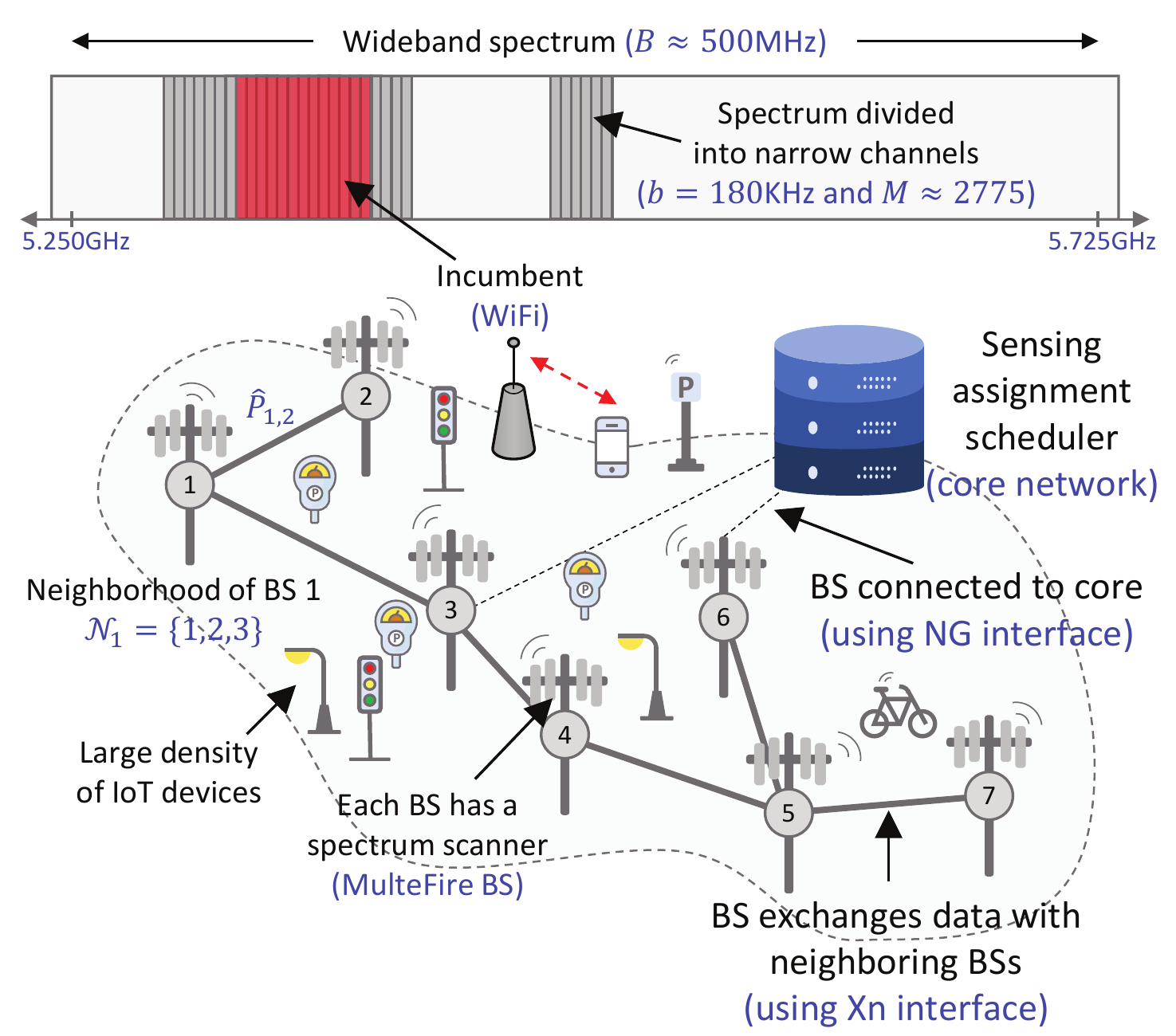}
	\caption{An illustration of the network architecture.}
	\label{fig:architecture}
\end{figure}

The output of the sensing algorithm provides BSs with a list of available channels in the wideband spectrum. If each BS then enables data exchange with IoT devices over any of these channels, collisions would occur within the IoT network as multiple nearby BSs could pick the same channel. Thus, we propose a distributed resource allocation algorithm that aims to solve the following problem: Given a set of available channels at each BS and its neighborhood, how should these spatio-spectral resources be allocated across BSs, i.e., reused over space, while limiting the intra-network and inter-network interference. An illustration of the network architecture is provided in Fig. \ref{fig:architecture}. 

%

\section{The Sensing Assignment Scheduler}\label{sec:scheduler}
In this section, we present an optimization framework for the sensing assignment problem and propose a low-complexity scheduler that meets the assignment constraints.

Let the set of channels assigned to the $k$-th BS be denoted as $\mathcal{M}_k\subset\mathcal{M}$. Since this BS will lack information about the channels in   $(\mathcal{M}_k)^c=\mathcal{M}\setminus \mathcal{M}_k$, it needs to collect sensing reports from neighboring BSs that sensed channels in $(\mathcal{M}_k)^c$. Nevertheless, collecting reports from distant BSs can be unreliable as the spatial footprints of incumbents vary over space. In other words, the sensing assignment scheduler should ensure that the distance between the $k$-th BS and any BS sensing channels in $(\mathcal{M}_k)^c$ is minimized. 

Let $c_{j,k,m}$ denote the cost of the $k$-th BS using the $j$-th BS sensing report of the $m$-th channel. For instance, the cost can be the quality of the received power of reference signals broadcasted by the $j$-th BS, or it can be the distance between the two BSs, which is independent of the channel. Further, let $\mathbf{X}\in\mathbb{Z}_2^{K\times M}$ be the assignment matrix, i.e., the $(k,m)$-th entry $x_{k,m}=1$ if the $k$-th BS is assigned to sense the $m$-th channel, and 0 otherwise. Then, we formulate the following sensing assignment problem\footnote{An underlying assumption here is $\mathbf{1}_M^T\mathbf{q}=\mathbf{1}_K^T\mathbf{p}$.}
\begin{equation}
\label{eq:IPOriginal}
\begin{array}{cl}
\underset{\mathbf{X}}{\text{{minimize}}}
&~~ \underset{m\in\{1,2\cdots,M\}}{\operatorname*{max}} \sum_{j=1}^K \sum_{k=1}^K c_{j,k,m} x_{k,m}\\\
\text{subject to}     
 &~~\mathbf{X}^T\mathbf{1}_K=\mathbf{q}, \\
&~~\mathbf{X}\mathbf{1}_M=\mathbf{p},\\
&~~\mathbf{X}\in \mathbb{Z}_2^{K\times M},
\end{array}
\end{equation}
where $\mathbf{q}=[q_1,q_2,\cdots,q_M]^T$ and $\mathbf{p}=[p_1,p_2,\cdots,p_K]^T$. 
In this formulation, the total cost of assigning a given number of BSs to sense an $m$-th channel is computed, and the objective is to minimize the maximum total costs across all channels, so that the $k$-th BS can reliably collect sensing reports for any channel in $(\mathcal{M}_k)^c$. The first constraint ensures that each $m$-th channel is sensed by $q_m$ BSs, and the second constraint ensures that each $k$-th BS senses only $p_k$ channels.
Since channels are narrowband, and incumbents like WiFi devices have wider bandwidth, then adjacent channels can have high frequency correlation, which can be further incorporated in the cost of sensing reports $c_{j,k,m}$ to encourage two neighboring BSs to sense adjacent channels. 

The problem in (\ref{eq:IPOriginal}) is a generalization of the bottleneck assignment problem, which is known to be NP-hard \cite{Martello1995}. Further, due to the large number of channels, $M$, it can be impractical to solve, and thus we consider a simpler framework as follows. First, we assume all BSs sense the same number of channels, i.e., $p_k=p\forall k$. In addition, we divide the spectrum into bands, instead of channels, where the number of bands is  $L=\lfloor\frac{B}{p\cdot b}\rfloor$. Therefore, each BS will sense a single band, which is a subset of $p$ consecutive channels, and each $l$-th band will be sensed by $\tilde q_l$ BSs. Thus, let $\tilde{\mathbf{X}}\in\mathbb{Z}^{K\times L}_2$ denote the sensing assignment matrix, with $\tilde x_{k,l}=1$ when the $k$-th BS is assigned the $l$-th band, and let $\tilde{c}_{j,k,l}$ be the cost of the $k$-th BS using the $j$-th BS sensing report of the $l$-th band. Then, we consider the following simpler integer program 
\begin{equation}
\label{eq:IPSimpler}
 \begin{array}{cl}
 \underset{\tilde{\mathbf{X}}}{\text{{minimize}}}
	&~~ \underset{l\in\{1,2\cdots,L\}}{\operatorname*{max}} \sum_{j=1}^K \sum_{k=1}^K \tilde c_{j,k,l} \tilde x_{k,l}\\\
 \text{subject to}     
&~~\tilde{\mathbf{X}}^T\mathbf{1}_K=\tilde{\mathbf{q}}, \\
&~~\tilde{\mathbf{X}}\mathbf{1}_L=\mathbf{1}_K,\\
&~~\tilde{\mathbf{X}}\in \mathbb{Z}_2^{K\times L},
\end{array}
\end{equation}
where the $(k,l)$-th entry of $\tilde{\mathbf{X}}$ is $\tilde x_{k,l}$ and $\tilde{\mathbf{q}}=[\tilde q_1,\tilde q_2,\cdots,\tilde q_L]^T$. The optimization problem in (\ref{eq:IPSimpler}) has a lower complexity than $(\ref{eq:IPOriginal})$ as $L\ll M$, and it is practical as each BS will sense a single block of $p$ channels instead of $p$ not-necessarily consecutive narrowband channels. However, it is still a combinatorial problem with high computational complexity when $K\gg1$. 

If we relax the integrality constraint in (\ref{eq:IPSimpler}), then the problem can be reformulated as a linear program (LP), where its optimal objective value is a lower bound on that of (\ref{eq:IPSimpler}), yet the optimal variables $\tilde x_{k,l}^\star$ are not necessarily integral. In addition, a simple rounding to $\tilde x_{k,l}^\star$ in this case may violate the assignment constraints. Thus, we propose the following rounding algorithm.

Let $\mathcal{I}$ denote the set of optimal variables, minimizing the linear program, that are integral, i.e., $\mathcal{I}=\{(k,l)|\tilde x_{k,l}^\star \in\{0,1\}\}$. Similarly, let $\mathcal{F}$ denote the set of optimal variables that are fractional. We define $\bar q_l=\sum_{(j,l)\in\mathcal{I}}  \tilde x_{j,l}^\star\forall l$, i.e., the number of BSs fully assigned to sense the $l$-th band. Note that if $\bar q_l=\tilde q_l\forall l$, then the optimal solution of the linear program is also optimal for (\ref{eq:IPSimpler}). Otherwise, if $\bar q_l<\tilde q_l$ for any $l$,  then we need to find $\tilde q_l - \bar q_l$ BSs to assign them to those bands. To do so, we first find the BS-band pair with the highest fractional value, i.e., $\tilde x_{k,l}^\star= \operatorname*{argmax}_{(j,m)\in\mathcal{F}} x_{j,m}^\star$. If we have for such pair $\bar q_l<\tilde q_l$, then we set $\tilde x_{k,l}^\star=1$ and $\tilde x_{k,m}^\star=0\forall m\neq l$, and we update $\mathcal{F}$. We repeat this process until $\bar q_l=\tilde q_l\forall l$. In essence, this algorithm starts with an incomplete assignment, yet with an objective value that is lower than the optimal value in (\ref{eq:IPOriginal}). Then, the algorithm keeps assigning the remaining BSs to the remaining bands, starting with the BS-band pairs with the highest optimized fractional variables. 

\subsection{A heuristic sensing assignment scheduler}
If we know the BSs' coordinates, we can directly use that information to perform sensing assignments because our objective is to ensure that for any BS $k$, every band in the spectrum is sensed by at least one BS in vicinity. Indeed, collecting reports from very far BSs should be discouraged as the spatial footprint of an incumbent varies over space. Thus, the set of BSs sensing the $l$-th band cannot be clustered in a particular area, but rather they should be spread out over the region. This can be accomplished, for instance, by dividing the region into several clusters, where within each one, the entire spectrum is sensed by its members. To this end, we propose the following algorithm.  

First, we pick a band, say $l$, and divide the BSs in the network into $\tilde q_l$ clusters $\mathcal{C}_{i,l}\subset{\mathcal{K}}\forall i=1,2,\cdots,\tilde q_l$, based on the coordinates of the BSs. From each cluster $\mathcal{C}_{i,l}$, we pick one BS that minimizes the worst cost, i.e., we solve the following problem for each cluster
\begin{equation}
\label{eq:IPCluster}
\begin{array}{cl}
\underset{\left\{\tilde x_{k,l}\in\mathcal{C}_{i,l};\tilde x_{k,l}\in\{0,1\}\right\}}{\text{{minimize}}}
&~~   \underset{j\in \mathcal{C}_{i,l}}{\operatorname*{max}}\sum_{k\in \mathcal{C}_{i,l}} \tilde c_{j,k,l} \tilde x_{k,l}\\\
\text{subject to}     
&~~\sum_{k\in\mathcal{C}_{i,l}} \tilde x_{k,l}=1, \\
\end{array}
\end{equation}
The optimal solution, in fact, is $\tilde x_{k,l}^\star=\{1|k=\operatorname*{argmin}_e  \sum_{j\in\mathcal{C}_{i,l}}\tilde c_{j,e,l}\}$. After this iteration, there remains $K-\tilde q_l$ BSs. Thus, we pick another band, say $u$, and then cluster the remaining BSs into $\tilde q_u$ clusters, and solve (\ref{eq:IPCluster}) for each cluster, repeating the process until all bands are completed. Note that the bands that are picked earlier in the procedure will have lower total cost as there are more BSs to pick from. To combat this, we repeat the whole process multiple times, randomizing the order of picked bands in each time. Then, we pick the one with the lowest maximum total cost.

\begin{algorithm}[!t] 	
	\small
	\caption{A heuristic sensing assignment scheduler}\label{alg:scheduler}
	\begin{algorithmic}[1]
		\Procedure{Input}{$\mathcal{K},\{\tilde c_{j,k,l}\},\tilde{\mathbf{q}},N$}
		\For{$n=1\longrightarrow N$}
		\State \textbf{Set} $\tilde{\mathcal{K}}=\mathcal{K}$		
		\State \textbf{Permutate} $\boldsymbol{l}=[1,2,\cdots,L]$
		\For{$u=1\longrightarrow L$}
		\State \textbf{Partition} $\tilde{\mathcal{K}}$ into $\tilde q_{l(u)}$ clusters $\{\mathcal{C}_{i,l(u)}\}_{i=1}^{\tilde q_{l(u)}}$
		\For{$i=1\longrightarrow \tilde q_{l(u)}$}
		\State \textbf{Solve} (\ref{eq:IPCluster}) to compute $\tilde x^\star_{k,l(u)}\forall k\in\mathcal{C}_{i,l(u)}$
		\EndFor
		\State \textbf{Update} $\tilde{\mathcal{K}}\rightarrow \tilde{\mathcal{K}}\setminus\{k|\tilde x_{k,l(u)}^\star=1\forall \mathcal{C}_{i,l(u)},i=1,\cdots,\tilde q_{l(u)}\}$ 
		\EndFor
		\State \textbf{Store} $\tilde{\mathbf{X}}^\star_n$ and its corresponding objective value $Z_n$
		\EndFor		
		\State \textbf{Return} $\tilde{\mathbf{X}}^\star_{n^\star}$, where $n^\star = \operatorname{argmin}_n Z_n$
		\EndProcedure
	\end{algorithmic}
\end{algorithm}

\subsection{Computational complexity and implementation of the heuristic scheduler}
The proposed algorithm is summarized in Alg. \ref{alg:scheduler}, and it has overall polynomial complexity. In particular, the algorithm computations are dominated by clustering the BSs. Using the $k$-means algorithm, clustering $K$ BSs into $q$ clusters based on their 2D coordinates  typically requires computations of order $O(2Kq)$ \cite{Bachem2016}. Comparing this with solving a linear relaxation of (\ref{eq:IPSimpler}), the latter requires computational complexity of order $O(K^3L^3)$ using interior-point methods, yet it does not require knowledge of BSs' coordinates. 

The assignment framework and proposed schedulers are centralized, i.e., they are solved at the core network. Thus, BSs need to forward the cost of collecting reports to the core network, or the BSs' coordinates may need to be known if the cost is the distance between BSs. The core network feeds back the assigned bands (and hence channels) to the BSs. Such two-way communication is done using existing interfaces, e.g., in 5G-NR, the NG interface connects the core network with BSs. We note that sensing assignments do not need to be frequently recomputed, as they primarily need to be updated when the network is changed, e.g., new BSs are deployed or existing ones are removed.

\subsection{Numerical Validation}
\begin{figure}[t!]
	\centering
	\begin{subfigure}[t]{.4\textwidth}
		\centering
		\includegraphics[width=3in]{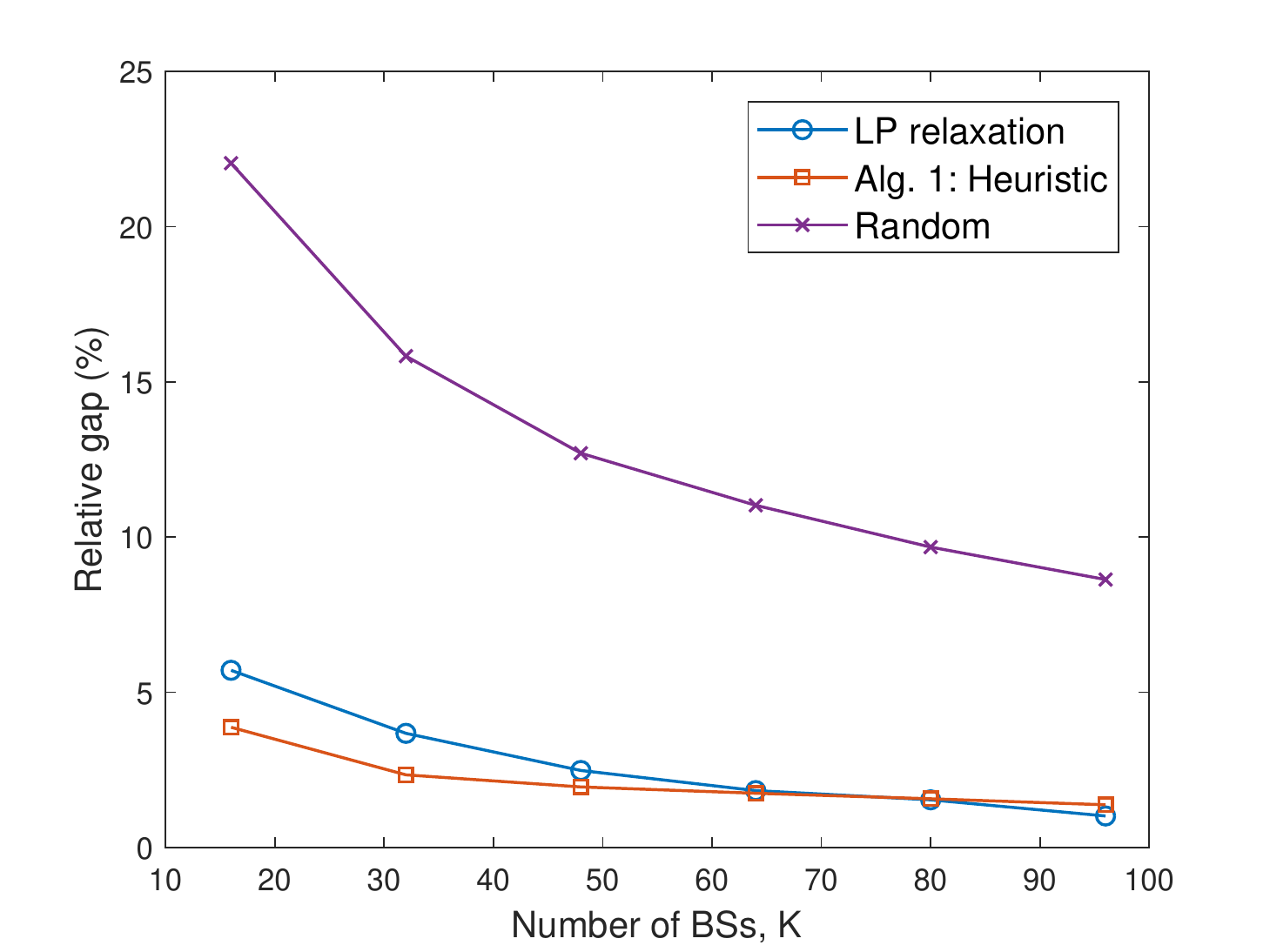}
		\caption{Relative gap versus $K$}
		\label{fig:relGapVsBSs}
	\end{subfigure}~
	\begin{subfigure}[t]{.4\textwidth}
		\centering
		\includegraphics[width=3in]{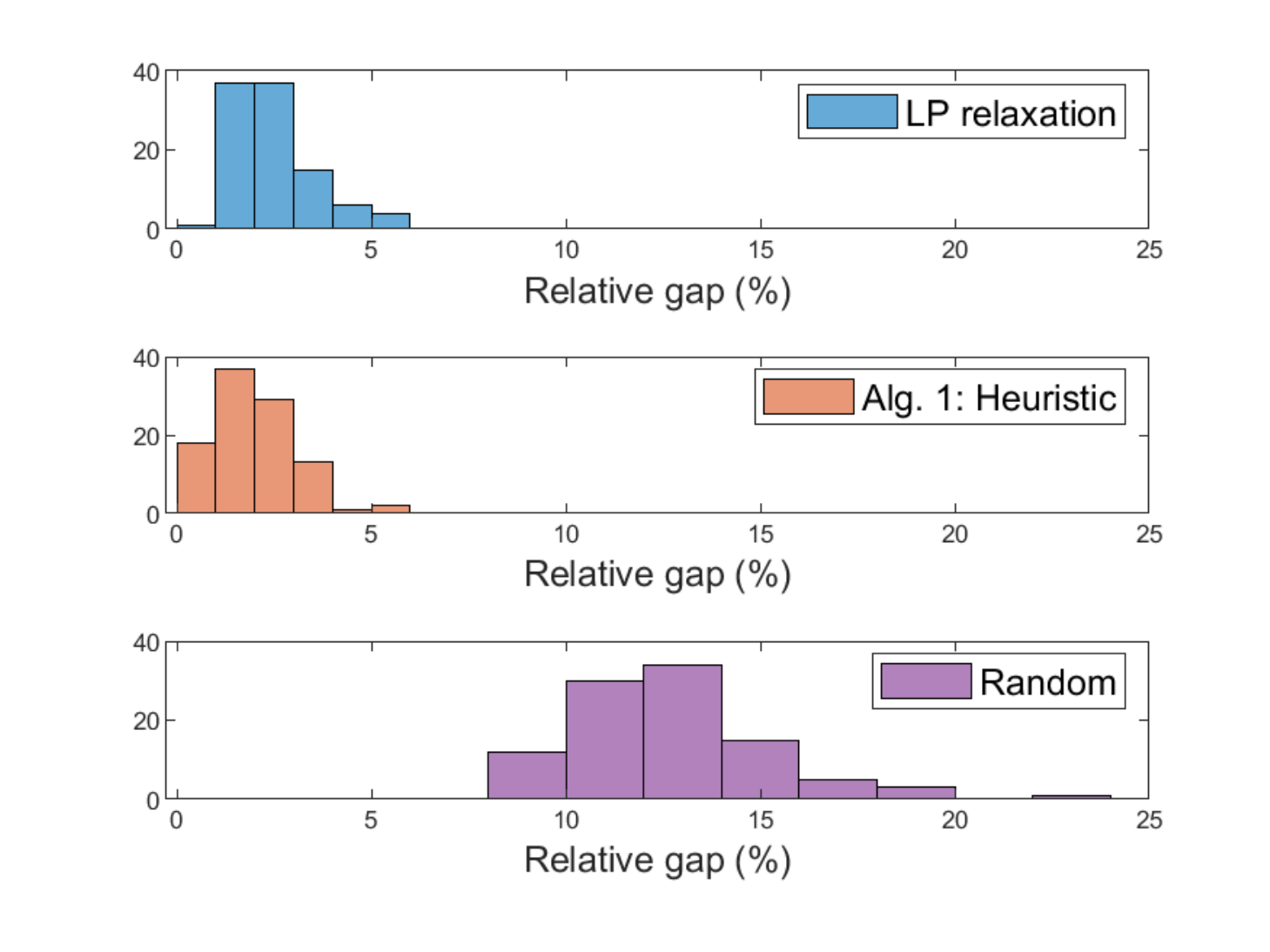}
		\caption{Histogram of relative gap ($K=48$)}
		\label{fig:relGapHist}
	\end{subfigure}
	\caption{Relative gap between the integer program (\ref{eq:IPSimpler}) and other scheduling algorithms.}
\end{figure}

We compare the solution of the integer program in (\ref{eq:IPSimpler}), which is solved  using MOSEK 8.0 solver with CVX in MATLAB, to the solution of the linear relaxation and Alg. \ref{alg:scheduler}. We also show the bottleneck performance  of random assignment that satisfies the assignment constraints. Due to the high complexity of solving the integer program, we only compare the solutions using relatively small number of BSs and bands. Specifically, we run 100 different realizations, where in each one we randomly deploy BSs on an area of $2\times2\text{km}^2$. We consider the costs to be equal to the distances between BSs. Here, we assume the number of bands to be $L=4$, and $\tilde{\mathbf{q}}=\tilde q\mathbf{1}$, and hence the number of BSs is $K=\tilde q\cdot L$. Fig. \ref{fig:relGapVsBSs} shows the average relative gap between the objective function in (\ref{eq:IPSimpler}) when evaluated at the optimal solution of the integer program and the solution obtained by the various assignment algorithms for different densities of BSs. It is observed that the proposed algorithms perform very well relative to the integer program, and the gap reduces for higher density of BSs, as higher density provides more flexibility of sensing assignments. We note that Alg. \ref{alg:scheduler} requires the coordinates of BSs, and such side information is useful, particularly when the density of BSs is lower. In Fig. \ref{fig:relGapHist}, we show the histogram of relative gaps across the different realizations when $K=48$. Unlike random assignment, the proposed algorithms achieve objective values within 5\% of the optimal integer program across the different realizations.  

\section{Diffusion-based Distributed Sensing}\label{sec:sensing}
In this section, we present the distributed wideband sensing algorithm and validate its performance relative to non-cooperative energy detection in terms of the deflection coefficient. 

Once sensing tasks are sent to BSs, each one measures the received energy in its assigned channels. For instance, in a MulteFire system, several CCA slots are used, where in each one, an energy measurement is compared to a detection threshold \cite{Kwon2017}. One major issue that can arise in such system is that raw energy estimates can considerably fluctuate over different CCA windows because of fading, shadowing, or blockage. To mitigate such abrupt changes, an adaptive least-mean-squares (LMS) filter is proposed in \cite{SobronVelez2015} so that the energy measurements are softened, and the filter weight is directly used as a test statistic. In particular, in LMS-based sensing, the BS minimizes the cost function $J(w_m)= \mathbb{E}[(d_{k,m}-w_m Y_{k,m})^2]$ by optimizing $w_m$, where it is desired to have $d_{k,m}= \mathbb{E}[Y_{k,m}]$ to reduce the variance across different energy measurements. Since the receiver does not have prior information about $\mathbb{E}[Y_{k,m}]$, $d_{k,m}$ is estimated in an online manner. The filter also admits a cooperative implementation, where the objective among cooperating BSs is to minimize $\bar J(w_m)= \sum_{k=1}^K \mathbb{E}[(d_{k,m}-w_m Y_{k,m})^2]$ \cite{SobronVelez2015}. 

While it is shown in  \cite{SobronVelez2015} that the proposed LMS-based approach significantly improves the detection performance in comparison with the energy detector, the approach requires all BSs to have the same $w_m$ to optimize. In other words, a cooperative procedure in this case implies that all BSs will aim to find the optimal estimate $w_m^\star$ that minimizes a global cost function, forcing each BS to have the same decision on the occupancy of the $m$-th channel, irrespective of its location. Since we aim to aggressively reuse channels over space, we need a fine-resolution frequency-space map of the spectrum. To this end, we propose to define a BS-specific cost function, i.e.,  $J_k(w_{k,m})= \mathbb{E}[(d_{k,m}-w_{k,m} Y_{k,m})^2]$, where the optimizing variable is BS-dependent. Although each BS has a different $w_{k,m}$ to estimate, nearby BSs that sense the $m$-th channel are expected to have correlated estimates due to the spatial correlation of energy footprints. Hence, cooperation among neighboring BSs can still improve the spectrum sensing reliability parallel to capturing the spatial variations of the incumbent's footprint. Thus, BSs cooperate to minimize $J_0(\boldsymbol{w}) = \sum_{k=1}^K J_k(\boldsymbol{w})$.

We aim to solve this global optimization distributively using diffusion-based distributed algorithms, which are known to be superior to other distributed strategies such as consensus and incremental strategies \cite{ChenSayed2015}. In this algorithm, each BS has a vector of weights $\boldsymbol{w}_{k}=[w_{k,1},w_{k,2},\cdots,w_{k,M}]^T$ to compute. Different from existing diffusion-based algorithms \cite{ChenSayed2015,Wang2017a}, the $k$-th BS only optimizes the vector entries corresponding to its assigned subset of channels, instead of optimizing all entries. The other entries will be  computed using the measurement reports of the BSs that sense different bands. 

The diffusion algorithm is centered around two main stages: the combination stage and the adaptation stage. 
In the combination stage, each BS shares its estimated $w_{k,m,i-1}$ (the subscript $i$ denotes the iteration number) with its neighbors, where we propose the following combination policy
\begin{equation}
\label{eq:combineStep}
\psi_{k,m,i-1} = \left\{
\begin{array}{ll}
 \sum_{j\in\mathcal{N}_{k}} \alpha_{jk,m,i} w_{j,m,i-1}, & m\in \mathcal{M}_k\\
 \sum_{j\in\mathcal{N}_{k}} \beta_{jk,i} w_{j,m,i-1}, &m\notin \mathcal{M}_k\\
\end{array}\right.,
\end{equation}
where $\{\alpha_{jk,m,i},\beta_{jk,i}\}$ are non-negative \emph{combining coefficients} that satisfy $\sum_{j\in\mathcal{N}_{k}} \alpha_{jk,m,i} =1$ and $\sum_{j\in\mathcal{N}_{k}} \beta_{jk,i}=1$. 
In the adaptation stage, the weights are updated as follows
\begin{equation}
\begin{aligned}
\label{eq:weight}
w_{k,m,i} &= \psi_{k,m,i-1} + \operatorname{{I}}_{(m\in\mathcal{M}_k)}\mu_{k} Y_{k,m,i}\left[d_{k,m,i}-Y_{k,m,i}\psi_{k,m,i-1}\right],
\end{aligned}
\end{equation}
where $\mu_{k}$ is a constant step-size and $d_{k,m,i}$ is an approximate of $\mathbb{E}[Y_{k,m}]$ \cite{SobronVelez2015}. In this work, we use an online first-order filter to compute $\mathbb{E}[Y_{k,m}]$. That is, we have 
\begin{equation}
\label{eq:d_ED}
d_{k,m,i}   = \zeta d_{k,m,i-1} + (1-\zeta) Y_{k,m,i},
\end{equation}
where $\zeta$ is a scalar close but less than one. It can be shown that $d_{k,m,i}$ has the same mean as $Y_{k,m}$ but with a lower variance (see the Appendix for the proof). 

Parallel to the aforementioned combination and adaptation stages, we perform an online clustering of BSs with similar weights by adapting  $\alpha_{jk,m,i}$. If the $k$-th BS is sensing the $m$-th channel, it assigns the following combining coefficient to the channel report sent from the neighboring $j$-th BS\cite{ChenSayed2015}
\begin{equation}
\label{eq:combinationweight_alpha}
\alpha_{jk,m,i} = \frac{(w_{k,m,i-1}+\mu_{k} \gamma_{k,m,i}-w_{j,m,i-1})^{-2}}{\sum_{j\in\mathcal{N}_{k}} (w_{k,m,i-1}+\mu_{k} \gamma_{k,m,i}-w_{j,m,i-1})^{-2}},
\end{equation}
where $\gamma_{k,m,i}=(d_{k,m,i}-Y_{k,m,i}w_{k,m,i-1})Y_{k,m,i}$. This weighting mechanism looks at the similarities between the estimated $w_{k,m,i-1}$ and $w_{j,m,i-1}$, where higher weight is given when these two values are closer to each other. Hence, as the algorithm progresses, cooperating BSs become clustered based on the similarities of their optimal solutions. If the $m$-th channel is not sensed by the $k$-th BS, then it will combine reports of this channel from neighboring BSs using the following coefficient
\begin{equation}
\label{eq:combinationweight_beta}
\beta_{jk} = \frac{\hat P_{k,j}}{\sum_{j\in\mathcal{N}_{k}\setminus \{k\}} \hat P_{k,j}},
\end{equation}
where we have  dropped the subscript $i$ as only one reference signal received power, per BS, is used. 

After $N$ iterations, each $k$-th BS will have $w_{k,m,N}\forall m\in\mathcal{M}$, which will be compared with a threshold $\lambda_{k,m}$ to make a decision on that channel. The threshold vector is computed by feeding the algorithm with energy threshold $\tau$, and using the output as a threshold for future samples. 

\begin{algorithm}[!t]
	\small
	\caption{Proposed distributed sensing algorithm implemented by the $k$-th BS}\label{alg:diffusion}
	\begin{algorithmic}[1]
		\Procedure{Input}{$\mu_{k},\mathcal{N}_{k}$,$\zeta$,$\hat P_{k,j}$}
		\For{$i=1\longrightarrow N$}
		\State \textbf{Measure} $Y_{k,m,i}$ and \textbf{Compute} $d_{k,m,i}   = \zeta d_{k,m,i-1} + (1-\zeta) Y_{k,m,i}$
		\State \textbf{Compute} $\gamma_{k,m,i}=(d_{k,m,i}-Y_{k,m,i}w_{k,m,i-1})Y_{k,m,i}$ 
		\State \textbf{Compute weights} $\alpha_{jk,m,i}$ using (\ref{eq:combinationweight_alpha}) and $\beta_{jk}$ using (\ref{eq:combinationweight_beta}) 
		\State \textbf{Combine} $\psi_{k,m,i-1}$ using (\ref{eq:combineStep}) and \textbf{Adapt}  $w_{k,m,i}$ using (\ref{eq:weight})
		\EndFor
		\State \textbf{Compare} $w_{k,m,N} \gtrless \lambda_{k,m}\forall k,m$
		\EndProcedure     
	\end{algorithmic}
\end{algorithm}

\subsection{Computational complexity and implementation of the distributed sensing algorithm}
The proposed distributed sensing algorithm is given in Alg. \ref{alg:diffusion}. Its computational complexity is as follows. For each $k$-th BS, the combination stage in (\ref{eq:combineStep}) requires $M|\mathcal{N}_k|$ operations, i.e., multiplications and additions, whereas the adaptation stage in (\ref{eq:weight}) requires $2M|\mathcal{N}_k|$ operations. The adaptive combining coefficients require operations of order $2M|\mathcal{N}_k|$. Thus, the overall computational complexity of the algorithm, per iteration, is $O(M|\mathcal{N}_k|)$. We emphasize that since each BS does not entirely sense the wideband spectrum, the sampling rate can be reduced, mitigating the need for high-rate analog-to-digital converters (ADCs). Alternatively, for the same sampling rate used for the wideband spectrum, the BS acquires more samples when it senses a subset of the spectrum, enabling finer spectral resolution and more sensing measurements.

The algorithm inevitably requires data exchange, yet it is restricted within the neighborhood of a BS. Thus, the number of data exchange messages transmitted by the BS in each iteration is proportional to $|\mathcal{N}_k|$. In each data exchange, the BS only shares the weights. Data communication between BSs can be done using standard interfaces, e.g., in 5G-NR, the Xn interface is used to connect two BSs using for example wired connections. In addition, control data exchange between the BS and IoT devices can be done using existing methods. For example, In LTE-LAA, a primary licensed carrier is used to carry control information, whereas in  MulteFire, discovery reference signals (DRS) are periodically sent \cite{MulteFire2017}, where upon decoding, the IoT device can learn which resource blocks have been allocated to it.

\subsection{Performance analysis of the deflection coefficient}
One metric that can be used to quantify the sensing performance of a test statistic $T$ is the deflection coefficient, which is expressed as
\begin{equation}
\label{eq:deflectionCoefficient}
\delta^2(T)= \frac{\left(\mathbb{E}_0[T]-\mathbb{E}_1[T]\right)^2}{\mathbb{V}_0(T)},
\end{equation}
where the subscripts $0$ and $1$ denote the absence or presence of incumbents, respectively. In essence, sensing is reliable when the distance in the mean between the test statistic under the presence and absence of incumbent is large, or when the variations of the test statistic is low. 

It is intractable to theoretically compute $\delta^2(w_{k,m})$, particularly because we use adaptive combining weights and realistic channel models. For this reason, we first use Monte Carlo simulations to show that Alg. \ref{alg:diffusion} improves the deflection coefficient under realistic channels. Then, we consider a simpler model and a special case of the proposed algorithm to theoretically compute the deflection coefficient.

\subsubsection{Numerical validation of Alg. \ref{alg:diffusion}}

\begin{figure}[t!]
	\centering
	\begin{subfigure}[t]{.32\textwidth}
		\centering
		\includegraphics[width=2.6in]{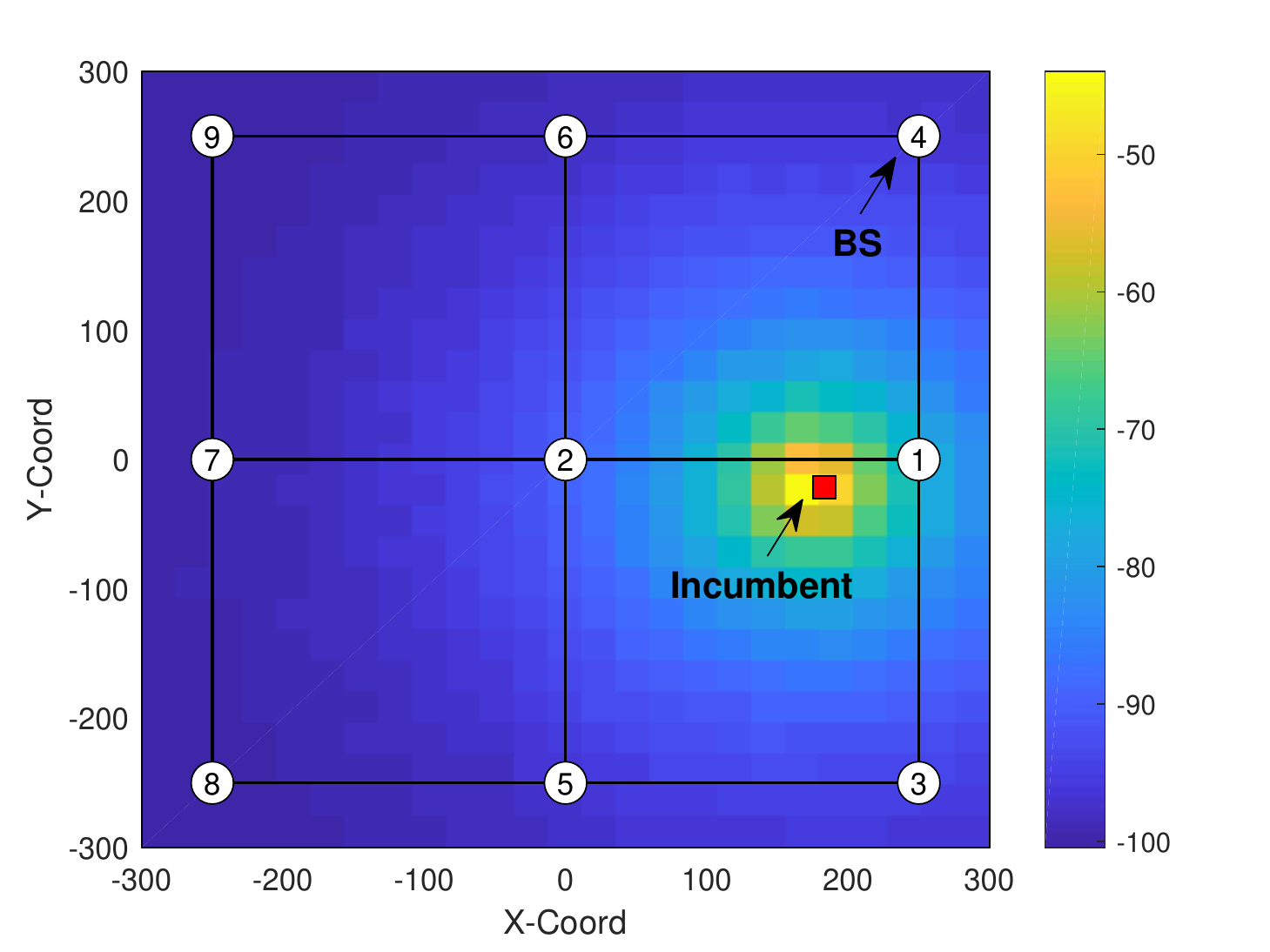}
		\caption{Incumbent's energy footprint}
		\label{fig:deflectionScenario}
	\end{subfigure}~~
	\begin{subfigure}[t]{.32\textwidth}
		\centering
		\includegraphics[width=2.6in]{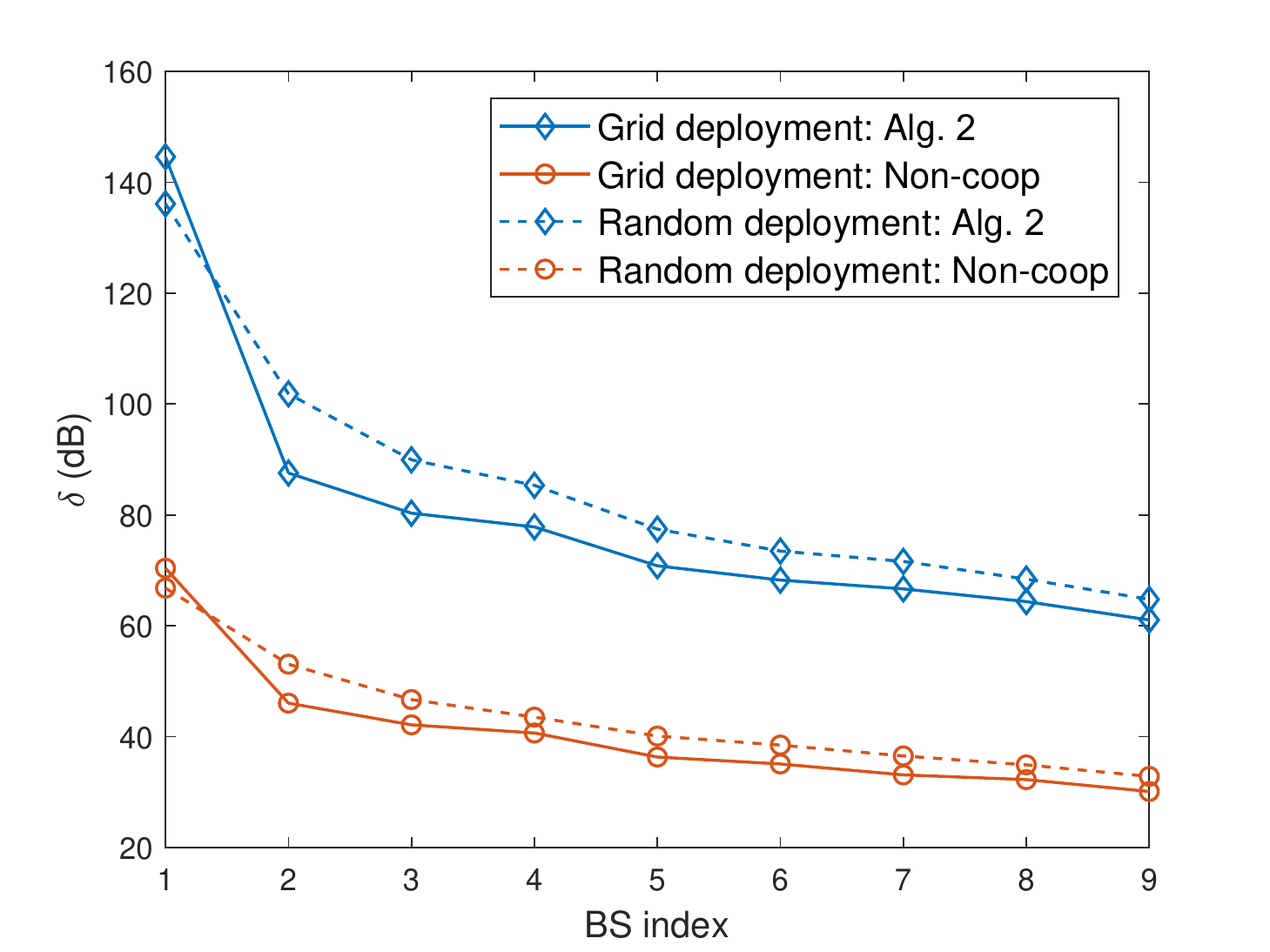}
		\caption{Deflection coefficient at different BSs}
		\label{fig:deflectionPerformance}
	\end{subfigure}~~
	\begin{subfigure}[t]{.32\textwidth}
		\centering
		\includegraphics[width=2.6in]{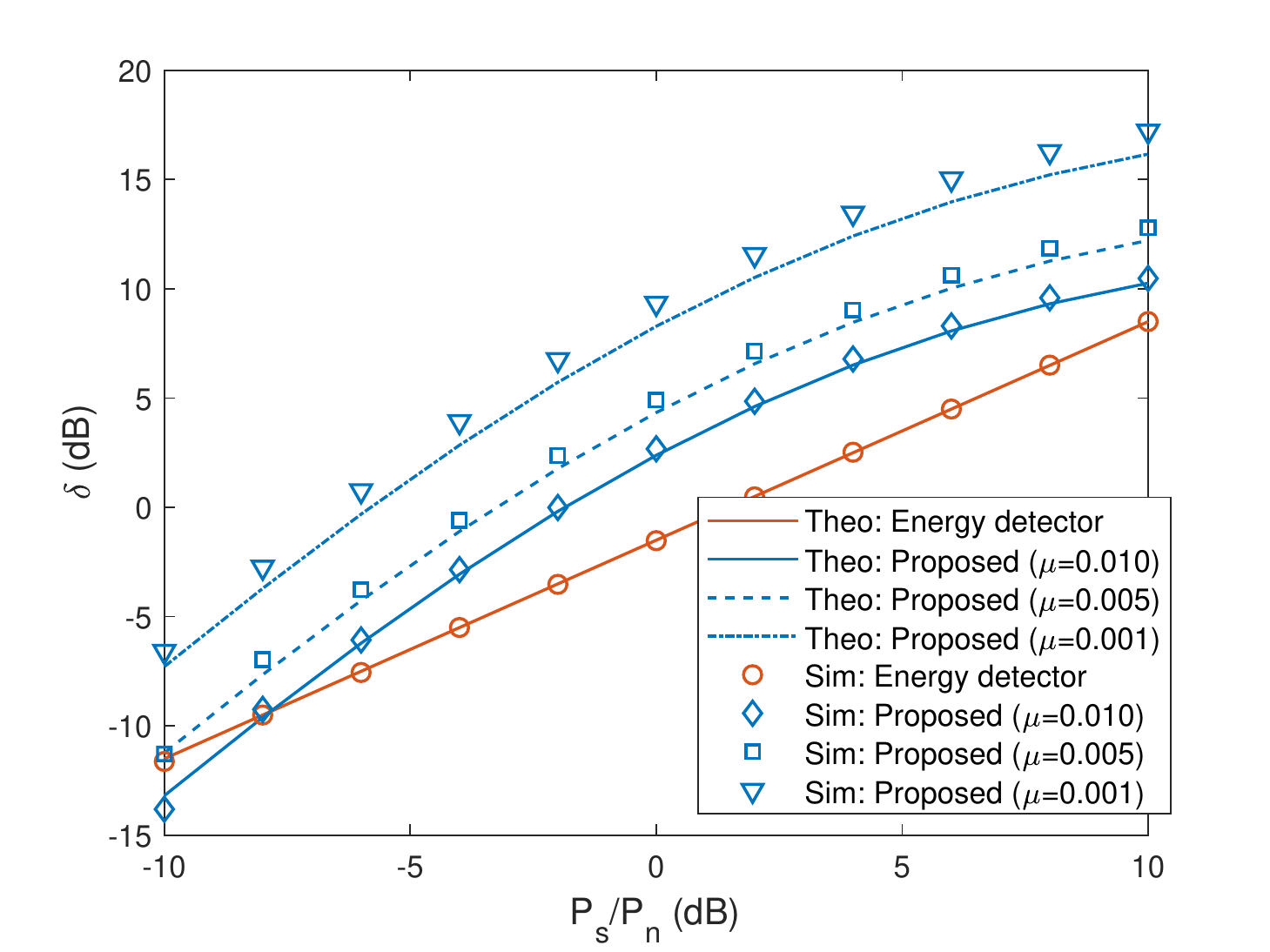}
		\caption{Deflection coefficient versus SNR}
		\label{fig:theorticalDeflection}
	\end{subfigure}
	\caption{Deflection coefficient comparison between the proposed and the non-cooperative sensing algorithms.}
\end{figure}

We consider a small network of $K=9$ BSs with two different deployments: a grid deployment and a random deployment. We run 1000 realizations, where in each one, we randomly drop a single incumbent that transmits at power $23$dBm when it is active. We then compute the deflection coefficients $\delta(Y_{k,m})$ and $\delta(w_{k,m})$ at each BS, which are indexed from the nearest to the furthest from the incumbent. It is assumed $\mu_k=0.01\forall k$. In Fig. \ref{fig:deflectionScenario}, we show one realization of the incumbent footprint when BSs are deployed on a grid. Here, the channel model is the outdoor 3GPP NR-UMi \cite[Sec. 7.4]{3GPP2017d}. We also show the connections between neighboring BSs. In Fig. \ref{fig:deflectionPerformance}, we show the average deflection at each BS. It is evident that all BSs, whether they are close or far from the incumbent, have superior deflection coefficients under the proposed sensing algorithm in comparison with the non-cooperative system.

\subsubsection{Theoretical analysis of the deflection coefficient}
Only in what follows, we make the following assumptions for tractable analysis. Specifically, we assume all BSs sense the same channel, have the same objective to minimize, and use non-adaptive combining coefficients $\alpha_{jk}$ which we collect in a matrix $\mathbf{A}$. Under this model, and with slight abuse of notation, we can rewrite (\ref{eq:combineStep}) in a matrix form as $\boldsymbol \psi_i = \mathbf{A}^T \boldsymbol w_i$, where $\boldsymbol w_i=[w_{1,i},w_{2,i},\cdots,w_{K,i}]^T$, i.e., the subscript $i$ here denotes the iteration index, and we drop the subscript $m$ in what follows because all BSs sense the same channel. Similarly, the adaptation stage in (\ref{eq:weight}) can be rewritten as 
\begin{equation}
\label{eq:adaptVector}
\boldsymbol w_i= (\mathbf{I}-\mathbf{M}\mathbf{Y}_i)\mathbf{A}^T \boldsymbol w_{i-1} + \mathbf{M} \boldsymbol{\phi}_i,
\end{equation}
where $\mathbf{M}=\operatorname{diag}(\mu_1,\mu_2,\cdots,\mu_K)$, $\mathbf{Y}_i=\operatorname{diag}(Y_{1,i}^2,Y_{2,i}^2,\cdots, Y_{K,i}^2)$, and $\boldsymbol{ \phi}_i=[Y_{1,i}d_{1,i},Y_{2,i}d_{2,i},\cdots,Y_{K,i}d_{K,i}]^T$.

In the steady-state, we have $\boldsymbol w_i \approx \boldsymbol w_{i-1}$, and thus taking the expectation on both sides of (\ref{eq:adaptVector}), we get 
\begin{equation}
\label{eq:meanEquation}
\mathbb{E}_{\kappa} [\boldsymbol w_i] = \left(\mathbf{I}-(\mathbf{I}-\mathbf{M} \mathbb{E}_{\kappa}[\mathbf{Y}_i])\mathbf{A}^T\right)^{-1} \mathbf{M}  \mathbb{E}_{\kappa}[\boldsymbol{\phi}_i],
\end{equation}
where $\kappa\in\{0,1\}$. To evaluate the variance $\mathbb{V}_0(w_{k,i})$, let $\tilde{\boldsymbol{w}_i}=\boldsymbol w_i- \mathbb{E}_{0} [\boldsymbol w_i]$. Since the objective is the same across BSs, we have $\mathbb{V}_{0}(w_{k,i})\approx \mathbb{E}_{0} [\|\tilde{\boldsymbol{w}_i}\|^2]/K$ \cite{Sayed2014a}. Let $\mathbf{B}=\mathbb{E}_0[(\mathbf{I}-\mathbf{M}\mathbf{Y}_i)\mathbf{A}^T]$, and let $\boldsymbol{\Sigma}$ be a nonnegative-definite matrix that we are free to choose. 
By following the energy-conservation argument \cite{Sayed2014a}, we can show that
\begin{equation}
\label{eq:EnergyConservation}
\mathbb{E}_0[\tilde{\boldsymbol{w}}_i^T(\boldsymbol{\Sigma}-\mathbf{B}\boldsymbol{\Sigma} \mathbf{B}^T)\tilde{\boldsymbol{w}}_i] =  \mathbb{E}_0[(\boldsymbol{\phi}_i-\mathbf{Y}_i \bar{\boldsymbol{w}_i})^T\mathbf{M}\boldsymbol{\Sigma}\mathbf{M}(\boldsymbol{\phi}_i-\mathbf{Y}_i \bar{\boldsymbol{w}_i})],
\end{equation}
where $\bar{\boldsymbol{w}_i}=\mathbb{E}_0[\boldsymbol{w}_i]$. By solving the discrete-time Lyapunov equation $\boldsymbol{\Sigma}-\mathbf{B}\boldsymbol{\Sigma} \mathbf{B}^T=\mathbf{I}$, we obtain $\boldsymbol{\Sigma}$. In this case, the left-hand side in (\ref{eq:EnergyConservation}) becomes $\mathbb{E}_{0} [\|\tilde{\boldsymbol{w}_i}\|^2]$, and thus the variance can be directly computed.

\subsubsection{Validation of the theoretical analysis}
Assume that the energy sample when the incumbent is absent is the energy of a Gaussian noise sample with variance $P_n$, and when the incumbent is present, it is the energy of a non-zero mean Gaussian sample with the same variance and with mean $\sqrt{P_s}$. Then, it can be shown that the deflection coefficient of the energy detector is $\delta(Y_k)=\frac{P_s}{\sqrt{2}P_n}$ \cite{SobronVelez2015}, where $P_s/P_n$ can be interpreted as the signal-to-noise ratio (SNR). We can also obtain simpler expressions for (\ref{eq:meanEquation}) and (\ref{eq:EnergyConservation}) under such assumptions (see the Appendix). In Fig. \ref{fig:theorticalDeflection}, we show the theoretical and simulated deflection coefficient with variations of the SNR. Here, we consider the grid deployment with $K=9$ BSs. We assume the same step-size for all BSs (shown in Fig. \ref{fig:theorticalDeflection}), and we use the averaging rule for the combining coefficients \cite{Sayed2014a}, i.e., $\alpha_{jk}=1/|\mathcal{N}_k|$ if $j\in\mathcal{N}_k$. It is shown that the theoretical analysis is in agreement with Monte Carlo simulations. Further, Alg. \ref{alg:diffusion} improves $\delta$, and the improvements are higher when smaller step-sizes are used.

\subsection{The proposed sensing algorithm as a REM}
In this section, we show that the weights $w_{k,m}$ can be interpreted as a REM as they capture the variations of received energies over space. REMs can be useful to study the coverage of a network, to optimize spectrum utilization and management, and to improve coexistence of different networks \cite{PerezRomero2015,Khan2017}. 

\begin{figure}[t!]
	\centering
	\begin{subfigure}[t]{.4\textwidth}
		\centering
		\includegraphics[width=3in]{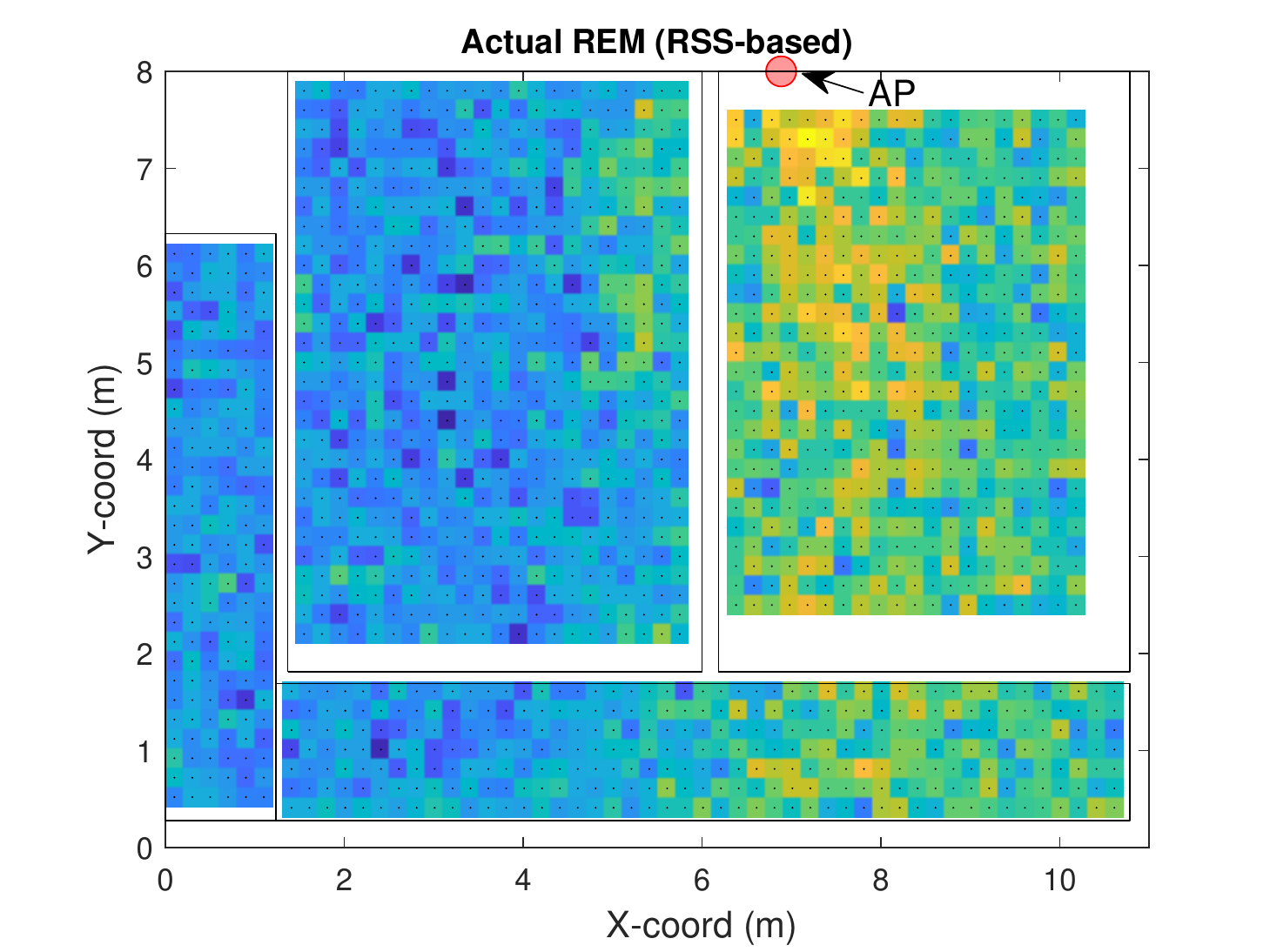}
		\caption{Actual REM}
		\label{fig:REM_RSS}
	\end{subfigure}~
	\begin{subfigure}[t]{.4\textwidth}
		\centering
		\includegraphics[width=3in]{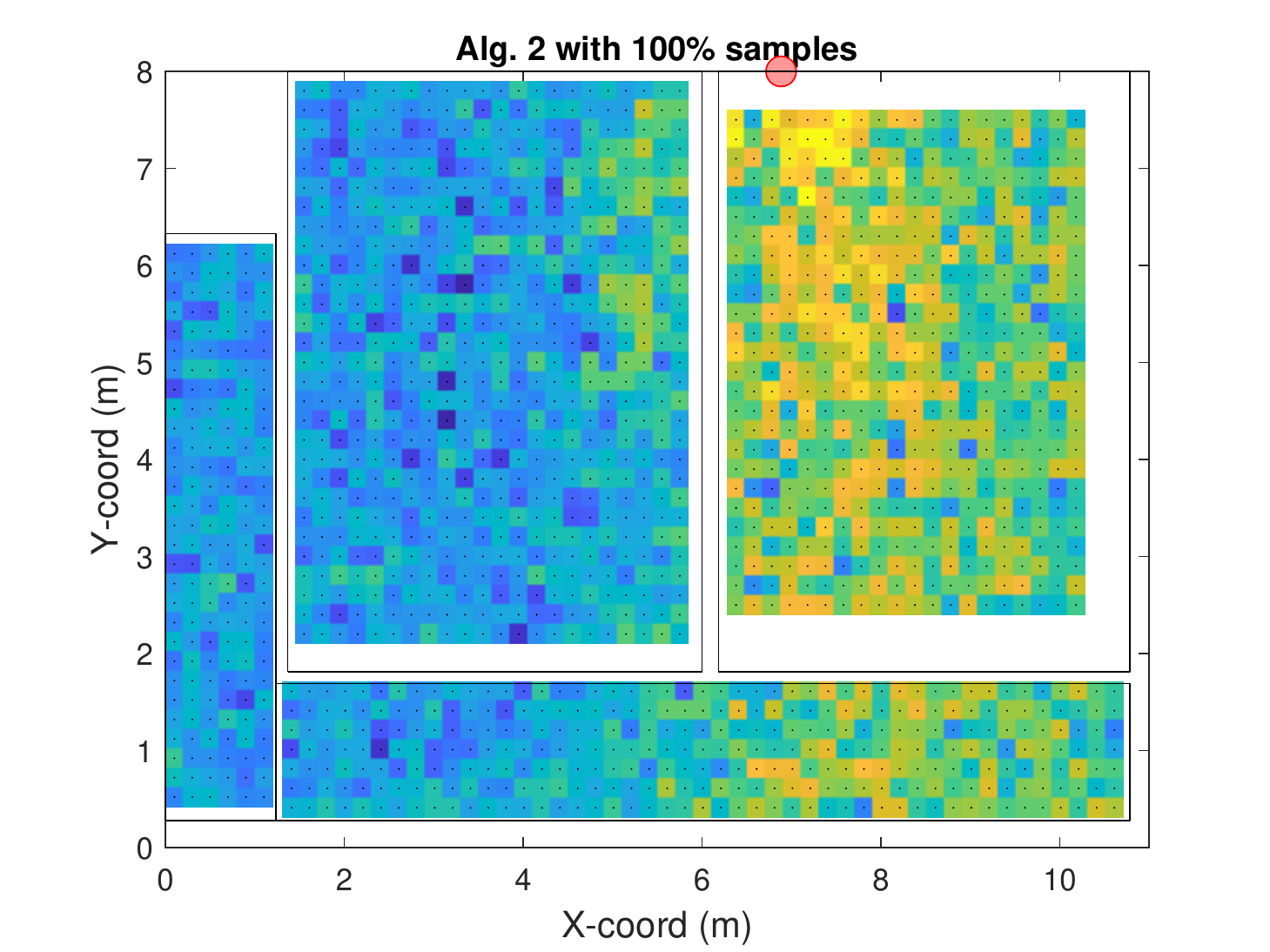}
		\caption{100\% of nodes using Alg. \ref{alg:diffusion}}
		\label{fig:REM_dist100}
	\end{subfigure}\\
	\begin{subfigure}[t]{.4\textwidth}
		\centering
		\includegraphics[width=3in]{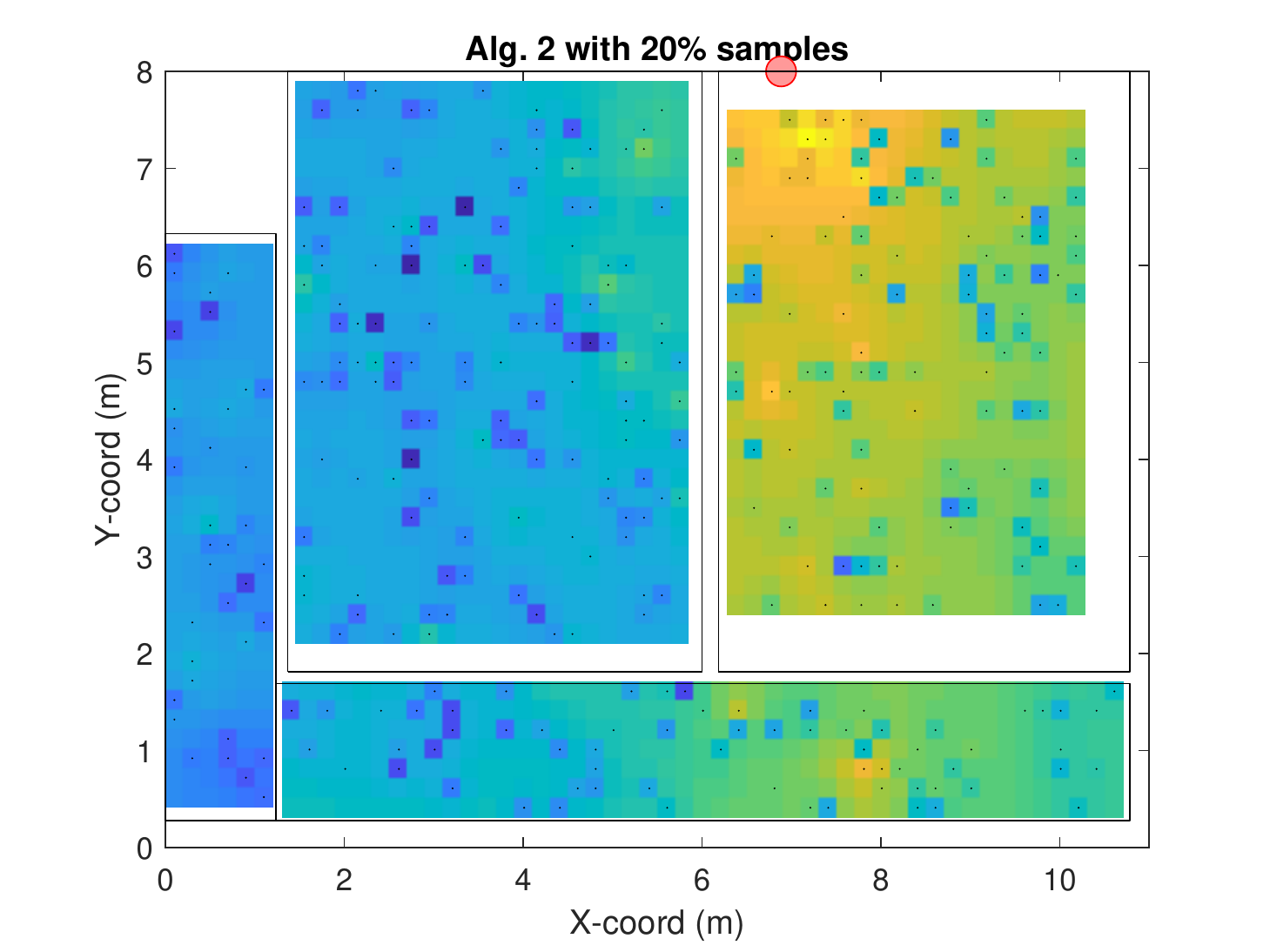}
		\caption{20\% of nodes using Alg. \ref{alg:diffusion}}
		\label{fig:REM_dist20}
	\end{subfigure}~
	\begin{subfigure}[t]{.4\textwidth}
		\centering
		\includegraphics[width=3in]{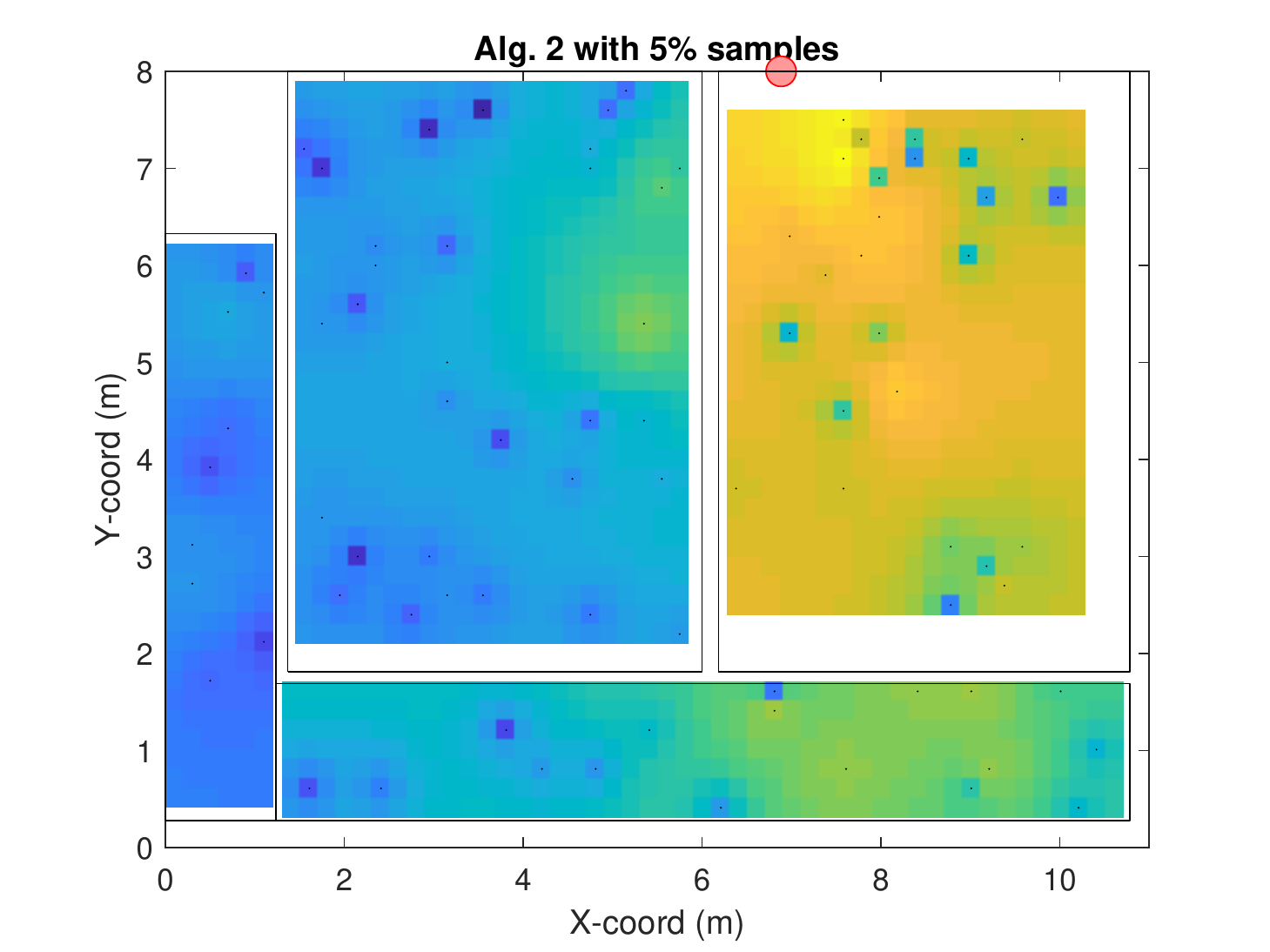}
		\caption{5\% of nodes using Alg. \ref{alg:diffusion}}
		\label{fig:REM_dist5}
	\end{subfigure}
	\caption{A REM for a single indoor WiFi AP using actual RSS measurements and the proposed algorithm.}
\end{figure}

To test the proposed algorithm on real data, we consider an office environment with one WiFi AP, using the set-up and data from \cite{Zakarya2018}. In the set-up, different sensing Raspberry PI boards (RPi3) are placed on the floor on a grid with inter-distance of 20cm. At each sampled location, henceforth denoted by a node, received signal strength (RSS) measurements are collected every three seconds, with a total of 40 RSS measurements per node.  Fig. \ref{fig:REM_RSS} shows the REM for this set-up. 

Next, we randomly select a subset of nodes, denoted by the set $\mathcal{S}$, and perform Alg. \ref{alg:diffusion} using the RSS measurements at the selected nodes. Each node cooperates with neighboring nodes that are within 2m, and obtain $w_{k}$ (we drop the subscript $m$). Then, we use the simple inverse distance weighting spatial interpolation (IDW) method \cite{Zakarya2018} at every $j$-th node in $(\mathcal{S})^c$, which is expressed as $w_{j} = \sum_{l\in\mathcal{S}} \beta_{lj} w_l\forall j\in(\mathcal{S})^c$, where here we set $\beta_{lj}$ to be inversely proportional to the distance squared between the $j$-th and $l$-th nodes \cite{Zakarya2018}. In Fig. \ref{fig:REM_dist100}, we show the output of the proposed algorithm when all nodes participate. It is evident that the weights, $w_k$, capture the same variations as the RSS measurements. In Fig. \ref{fig:REM_dist20} and  Fig. \ref{fig:REM_dist5}, we show the weights when 20\% and 5\% of the nodes participate and the rest use the IDW interpolation, respectively. Even with lower density of sensing nodes, the weights still provide a good estimate of the REM.

\section{Distributed Allocation of Spatio-spectral Resources}\label{sec:access}
In this section, we present a network utility optimization where the spatio-spectral blocks, identified by Alg. \ref{alg:diffusion}, are allocated across BSs for access. We then present a distributed allocation algorithm, where neighboring BSs locally allocate these blocks among them. 

\subsection{Network utility optimization}
If each BS uses all identified spectral opportunities for IoT access, then nearby BSs may interfere with each other. Thus, we propose that neighboring BSs do not use the same channel, where our objective is to allocate the spectral blocks, with the highest spatial reuse under the aforementioned constraint. 

We define the reward $r_{k,m}$ if the $k$-th BS is allowed to use the $m$-th channel. Since we do not rely on side information or feedback from the associated IoT devices, e.g., no CSI is available beforehand, we aim to merely use the outputs of Alg. \ref{alg:diffusion} to quantify the reward, i.e., $\{w_{k,m,N},\lambda_{k,m},d_{k,m,N}\}$.  To this end, we propose the following reward value $r_{k,m} = 10\log_{10}(\frac{\tau}{d_{k,m,N}})$.
The motivation behind this reward is twofold. First, the ratio $\frac{\tau}{d_{k,m,N}}$ quantifies how far the filtered energy measurement is  from the detection threshold. Indeed, a larger ratio implies that the estimated aggregate incumbents' received power is lower in the $m$-th channel, and thus using such channel is useful for both the IoT network and the incumbents. Second, the logarithmic function encourages fairness among BSs. Thus, we have the following optimization problem
\begin{equation}
\label{eq:centralizedAllocation}
\begin{array}{cl}
\underset{\{z_{k,m}\}}{\text{{maximize}}}
&~~ \sum_{m=1}^M \sum_{k=1}^K r_{k,m} z_{k,m}\\\
\text{subject to}     
&~~\sum_{j\in\mathcal{N}_k} z_{j,m}\leq 1, ~~\forall k,m\\
&~~z_{k,m} \leq \operatorname{{I}}_{(w_{k,m,N}\leq \lambda_{k,m})},  ~~\forall k,m\\
&~~ z_{k,m}\in\{0,1\},\\
\end{array}
\end{equation}
where $z_{k,m}$ is the optimizing variable, i.e., $z_{k,m}=1$ when the $k$-th BS is allocated the $m$-th channel for access. The first constraint ensures that within the neighborhood of a BS, a channel is used by at most one BS to limit the intra- and inter-network interference. The second constraint ensures that a BS cannot be assigned a channel that is deemed unavailable by the sensing algorithm. 

The integer program in (\ref{eq:centralizedAllocation}) can be solved in a distributed manner using the dual decomposition method. In particular, it is observed that the problem can be decomposed into $M$ separable problems, one per channel. For each $m$-th problem, let $\mathbf{z}_{m}=[z_{1,m},z_{2,m},\cdots,z_{K,m}]^T$, and introduce the dual variables $\boldsymbol \nu_{m}\geq0$ to decouple the first constraint in (\ref{eq:centralizedAllocation}). Thus, the dual problem for the $m$-th channel can be shown to be 
\begin{equation}
\label{eq:dual}
\underset{\boldsymbol \nu_{m} \ge 0}{\text{{minimize}}}  ~~ \mathbf{1}_{K}^T\boldsymbol \nu_{m} +  \sum_{k=1}^K D_k(\mathbf{z}_m,\boldsymbol \nu_{m}),
\end{equation}
where $D_k(\mathbf{z}_m,\boldsymbol \nu_{m})$ is the Lagrangian at the $k$-th BS, which is expressed as
\begin{equation}
\label{eq:Lagrangian}
D_k(\mathbf{z}_m,\boldsymbol \nu_{m}) = ~~\operatorname*{minimize}_{0\leq z_{k,m}\leq \operatorname{{I}}_{(w_{k,m,N}\leq \lambda_{k,m})}} z_{k,m}(r_{k,m} - \sum_{j\in\mathcal{N}_k}\nu_{j,m}).
\end{equation} 
If we fix the dual variables, then the solution to (\ref{eq:Lagrangian}) is 
\begin{equation}
\label{eq:primalStep}
z_{k,m}^\star = \operatorname{{I}}_{(w_{k,m,N}\leq \lambda_{k,m})} \times \operatorname{{I}}_{(r_{k,m} \geq \sum_{j\in\mathcal{N}_k}\nu_{j,m})}.
\end{equation} 
Similarly, since (\ref{eq:dual}) is differentiable, we can solve it using the gradient method, for a given $\mathbf{z}_m$. This process is recursively done until the dual converges. To summarize, the dual decomposition admits the following distributed implementation during the $i$-th iteration:
\begin{itemize}
	\item The selection step: Each BS first collects $\{\nu_{j,m,i}\}$ from its neighboring BSs, and computes (\ref{eq:primalStep}) for each channel ,i.e., the BS selects an available $m$-th channel if there is a non-negative profit defined by $r_{k,m}-\sum_{j\in\mathcal{N}_k}\nu_{j,m,i}$.
	\item The pricing step: Each BS collects the allocation decisions of neighboring BSs, and then solves the dual problem using the gradient method, i.e., each BS computes
	\begin{equation}
	\label{eq:priceUpdate}
	\nu_{k,m,i+1} = \operatorname{max}\left\{\nu_{k,m,i} - \tilde \mu_{k,i} (1-\sum_{j\in\mathcal{N}_k} z^\star_{j,m}),0\right\},
	\end{equation}
	where $\tilde \mu_{k,i}$ is a predetermined step-size. 
\end{itemize}
The dual variable can be interpreted as a price of selecting a channel. Indeed, if $\sum_{j\in\mathcal{N}_k} z^\star_{j,m}>1$ in (\ref{eq:priceUpdate}) then the price consistently increases in each iteration until only one BS can afford that channel. 
Although the dual decomposition algorithm is distributed, the performance of the algorithm requires to carefully select or update the step-size, and gradient-based methods can have slow convergence. 

\subsection{Fast distributed resource allocation}
In this section, we present a heuristic, yet fast distributed resource allocation algorithm. The key idea is to avoid incremental increases of prices that are needed so only a single BS can afford the current price within its neighborhood. To do so, we define the highest possible reward of selecting an $m$-th channel in a given neighborhood, i.e., $\bar r_{k,m,0} = \operatorname{max}_{j\in\mathcal{N}_k} r_{j,m}$. Then, each BS initially  selects the $m$-th channel if its reward $r_{k,m}$ is at least as large as $\bar r_{k,m,0}$. The selection stage is followed by a conflict resolution stage, which is resolved at every BS that has a channel selected within its neighborhood. Such conflict will be resolved by propagating the highest possible rewards among neighboring BSs. In particular, at the $i$-th message exchange, the $k$-th BS checks whether a channel is selected by any member in $\mathcal{N}_k$. If $\sum_{j\in\mathcal{N}_k}z_{j,m}\geq 1$, the BS recomputes the possible reward over the channel as follows
\begin{equation}
\bar r_{k,m,i+1}  = \operatorname*{max}_{j\in\mathcal{N}_k} \{z_{j,m} \bar r_{j,m,i}\}.
\end{equation}
The BS then gives up the channel in the next iteration if the maximum among the new rewards exceeds its own reward, i.e., $r_{k,m} < \operatorname{max}_{j\in\widetilde{\mathcal{N}}_k} \{\bar r_{j,m,i},r_{j,m}\}$, where $\widetilde{\mathcal{N}}_k=\{j|j\in\mathcal{N}_k,\sum_{l\in\mathcal{N}_j}z_{l,m}\geq 1\}$ is the set of BSs that participate in the conflict resolution process. A key observation here is that if BS $k$ gives up a channel that it initially selected, then in the next iteration, we must have $k\notin\widetilde{\mathcal{N}}_j\forall j\in\mathcal{N}_k$. Thus, the neighbors of BS $k$ can now contend for that channel without worrying about the fact that BS $k$ had a higher reward. 

\begin{algorithm}[!t]
	\small
	\caption{Proposed fast distributed allocation implemented by the $k$-th BS for the $m$-th channel}\label{alg:allocation}
	\begin{algorithmic}[1]
		\Procedure{Input}{$r_{k,m},\mathcal{N}_{k}$} 
		\State \textbf{Initialize}  $\bar r_{k,m,0} = \operatorname{max}_{j\in\mathcal{N}_k} r_{j,m}$ and $\widetilde{\mathcal{N}}_k=\mathcal{N}_k$
		\For{$i=1\longrightarrow N$}
		\State \textbf{Compute} $z_{k,m,i}= \operatorname{{I}}_{(w_{k,m,N}\leq \lambda_{k,m})} \times \operatorname{{I}}_{(r_{k,m}   \geq \operatorname{max}_{j\in\widetilde{\mathcal{N}}_k} \{\bar r_{j,m,i},r_{j,m}\})}$ 
		\If{$(\sum_{j\in\mathcal{N}_k}z_{j,m,i}\geq 1)$}
		\State \textbf{Update} $\bar r_{k,m,i+1}  = \operatorname*{max}_{j\in\mathcal{N}_k} \{z_{j,m,i} \bar r_{j,m,i}\}$
		\Else
		\State \textbf{Update} $\bar r_{k,m,i+1}=\bar r_{k,m,i}$
		\EndIf
		\State \textbf{Update} $\widetilde{\mathcal{N}}_k=\{j|j\in\mathcal{N}_k,\sum_{l\in\mathcal{N}_j}z_{l,m,i}\geq 1\}$ 
		\EndFor
		\EndProcedure     
	\end{algorithmic}
\end{algorithm}

\subsection{Implementation of the fast allocation algorithm}

The proposed fast distributed allocation is summarized in Alg. \ref{alg:allocation}. In terms of implementation, neighboring BSs need to share their original utility once. Then, in every iteration, the BS shares the maximum observed reward and whether it should be included in the conflict resolution process. In our simulations, the algorithm converges within a couple of iterations, and it performs relatively well compared to the centralized solution. 

\begin{figure}[t!]
	\centering
	\begin{subfigure}[t]{.4\textwidth}
		\centering
		\includegraphics[width=3in]{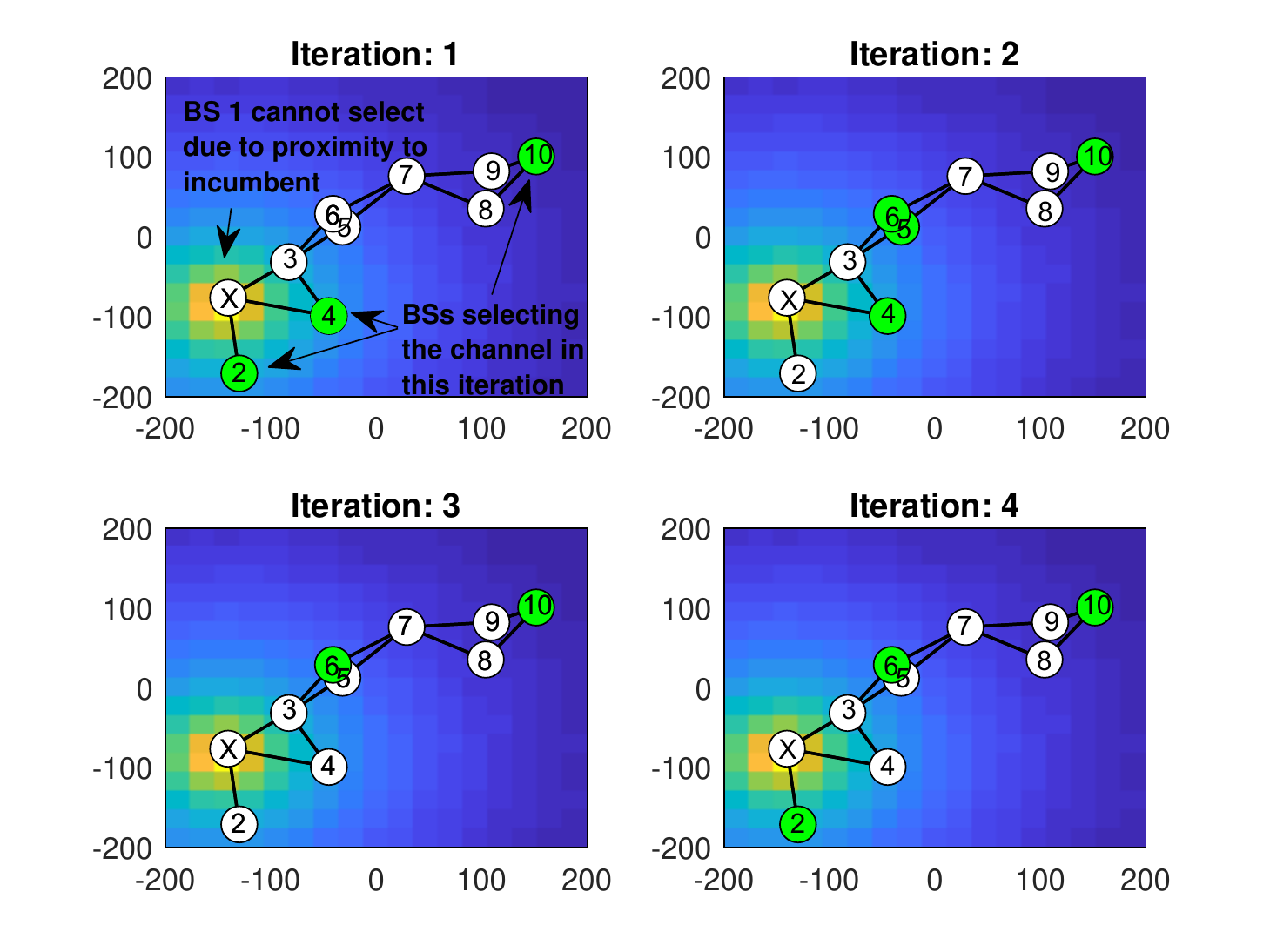}
		\caption{Distributed allocation over different iterations}
		\label{fig:Access_scenario}
	\end{subfigure}~
	\begin{subfigure}[t]{.4\textwidth}
		\centering
		\includegraphics[width=3in]{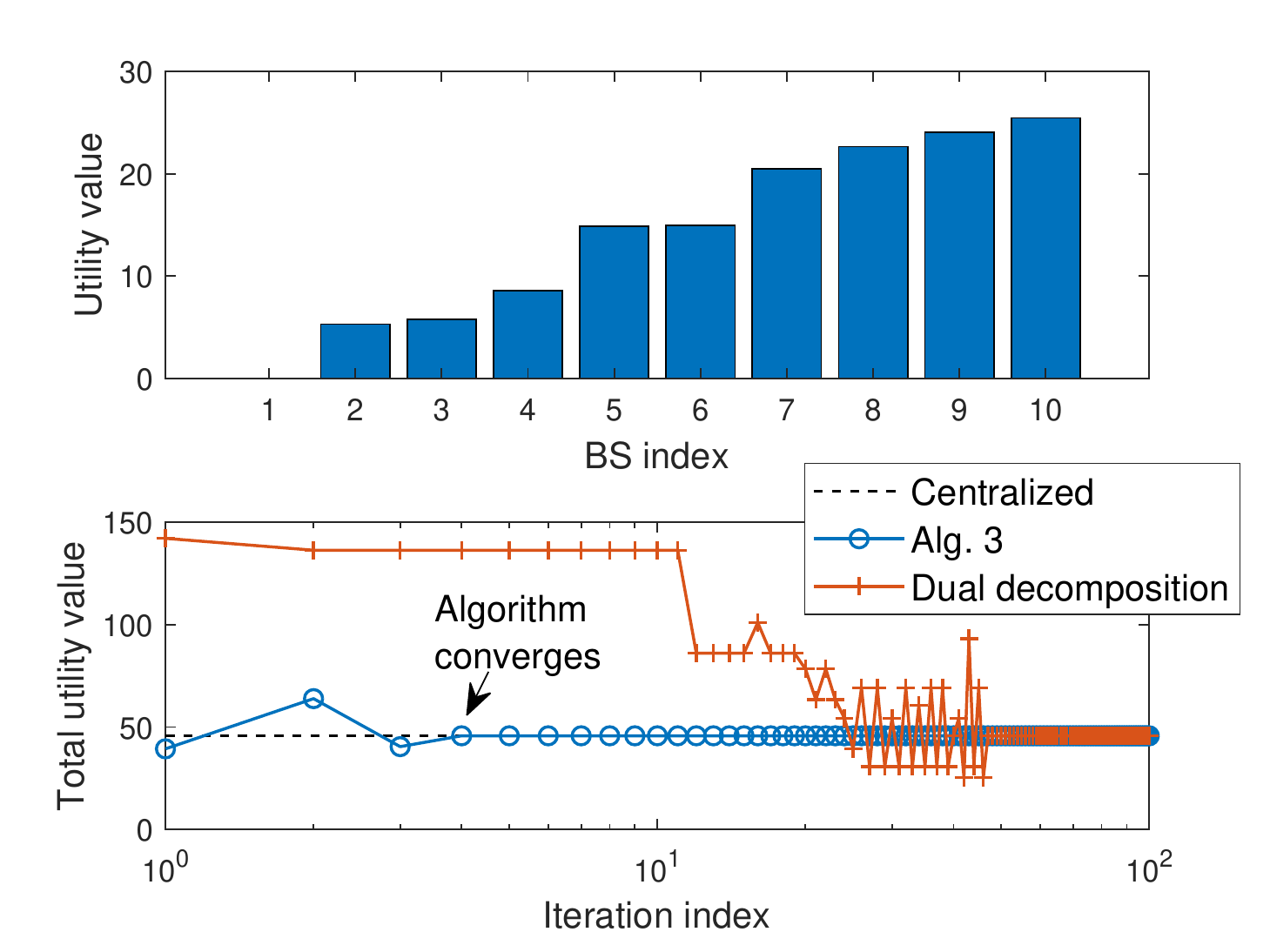}
		\caption{Utility over time}
		\label{fig:utility_vs_iteration}
	\end{subfigure}
	\caption{Illustration of distributed allocation over a small network.}
\end{figure}

For illustration purposes, we consider a small network with $K=10$ and a single channel with one active WiFi AP. In Fig. \ref{fig:Access_scenario}, we show the decisions made by BSs in each iteration. For example, in the first iteration, only BSs $\{2,4,10\}$ select the channel. However, since BS 1 now has two neighbors selecting the channel, a conflict resolution is used in the next iteration. In Fig. \ref{fig:utility_vs_iteration}, we show the utility of each BS, and the total utility over time when Alg. \ref{alg:allocation} and the dual decomposition are used. The former converges to the solution of the centralized scheme in four iterations, whereas the latter requires 47 iterations.

\begin{table}[!t]
	\caption{Simulation parameters}
	\label{tab:simulations}
	\centering
	\begin{tabular}{|l|c|c|}
		\hline
		Parameter   			&  Small-scale network & Large-scale network\\\hline
		BS deployment			& Grid				   &Random\\\hline
		No. of BSs  			& $K=100$ 			   &$K=500$\\\hline
		No. of incumbents		& 20				   &2000 WiFi APs\\\hline
		No. of channels			& $M=4$ 			   &$M=350$; $M=2775$\\\hline
		Spectrum bandwidth		& $B=80$MHz			   &$B=500$MHz\\\hline
		Channel bandwidth       & $b=20$MHz			   &$b=1.4$MHz; $b=180$KHz\\\hline
		Neighborhood radius		& $R=200$m			   & $R=1$km\\\hline
		Channel model		    & \multicolumn{2}{c|}{Outdoor 3GPP NR-UMi \cite{3GPP2017d}} \\\hline 
		Detection threshold		& \multicolumn{2}{c|}{$\tau=-62$dBm}\\\hline 
		Sensing algorithm 		& \multicolumn{2}{c|}{$\mu_k=0.01\forall k$ and $\zeta=0.95$}\\\hline 
	\end{tabular}
\end{table}

\section{Simulation Results}\label{sec:simulations}
We study the performance of the proposed architecture, comparing it with existing schemes. Specifically, the first two sections focus on a small-scale network, where the  distributed sensing and resource allocation algorithms are evaluated, respectively. The last section considers a large-scale case study with massive IoT used for smart public parks. Unless otherwise stated, Table \ref{tab:simulations} lists the values of the  parameters used.

\subsection{Performance evaluation of the distributed sensing algorithm}

\subsubsection{Set-up}
In this set-up, we randomly deploy 20 incumbent devices, where each one randomly picks one of the four channels, each of bandwidth 20MHz. All incumbents transmit at $23$dBm. For the channel model, we consider the 3GPP NR-UMi model, which is suitable for dense urban areas. We assume that the spectrum is centered around 5.43GHz, and all BSs and WiFi APs are deployed at a height of 10m. We run 100 realizations, where incumbents' locations, their channel usage, and propagation losses are randomized across realizations. We then consider the following schemes:
\begin{itemize}
\item \textbf{Distributed wideband}: We implement Alg. \ref{alg:diffusion}, yet assume all BSs sense the entire spectrum.
\item \textbf{Distributed narrowband}: We implement Alg. \ref{alg:diffusion}, where each BS senses a single channel that is assigned via Alg. \ref{alg:scheduler}. Similar results are obtained if the scheduler is based on the linear relaxation of (\ref{eq:IPSimpler}), and thus results are omitted. 
\item \textbf{Centralized}: We divide BSs into 25 clusters, where a cluster head collects energy measurements from the cluster members, and then combines them using equal gain combining, making one decision for the entire cluster. We only consider BSs with wideband sensors.
\item \textbf{Non-cooperative wideband}: Each BS senses the entire spectrum and makes a local decision about the availability of each channel. 
\item \textbf{Non-cooperative narrowband}: Each BS randomly picks a channel to sense. The BS will not have information about other channels. 
\end{itemize}  
We compare the aforementioned schemes in terms of the \emph{utilization ratio}, defined as the ratio of spatio-spectral blocks that are correctly identified as available by the scheme relative to those found by the genie scheme that knows all correct decisions, and in terms of the \emph{misdetection probability}, which is defined as the probability of incorrectly deciding a spatio-spectral block is available. 

\subsubsection{Impact of detection threshold}
\begin{figure}[t!]
	\centering
	\begin{subfigure}[t]{.4\textwidth}
		\centering
		\includegraphics[width=3in]{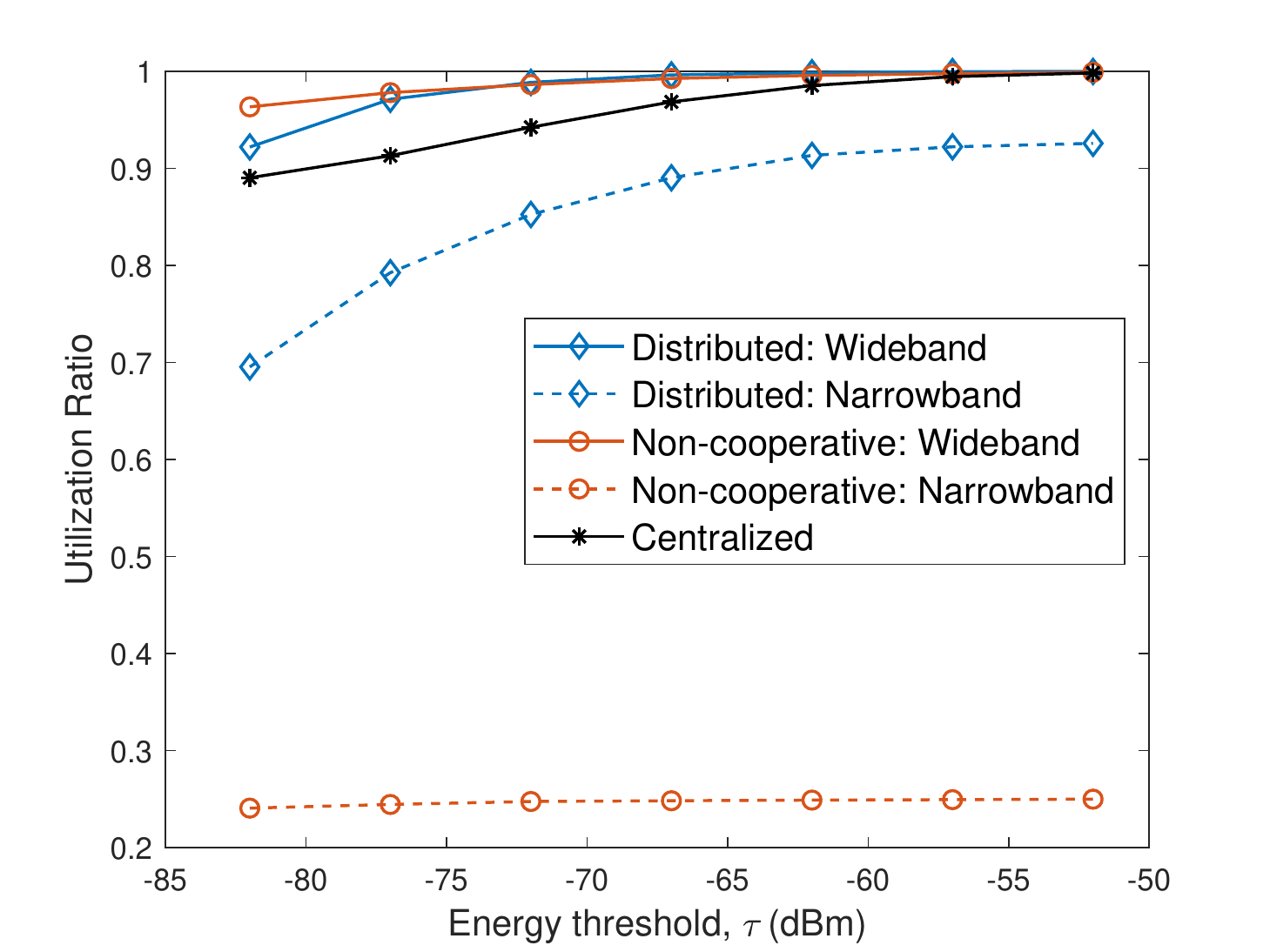}
		\caption{Utilization ratio}
		\label{fig:UT_vs_threshold}
	\end{subfigure}~
	\begin{subfigure}[t]{.4\textwidth}
		\centering
		\includegraphics[width=3in]{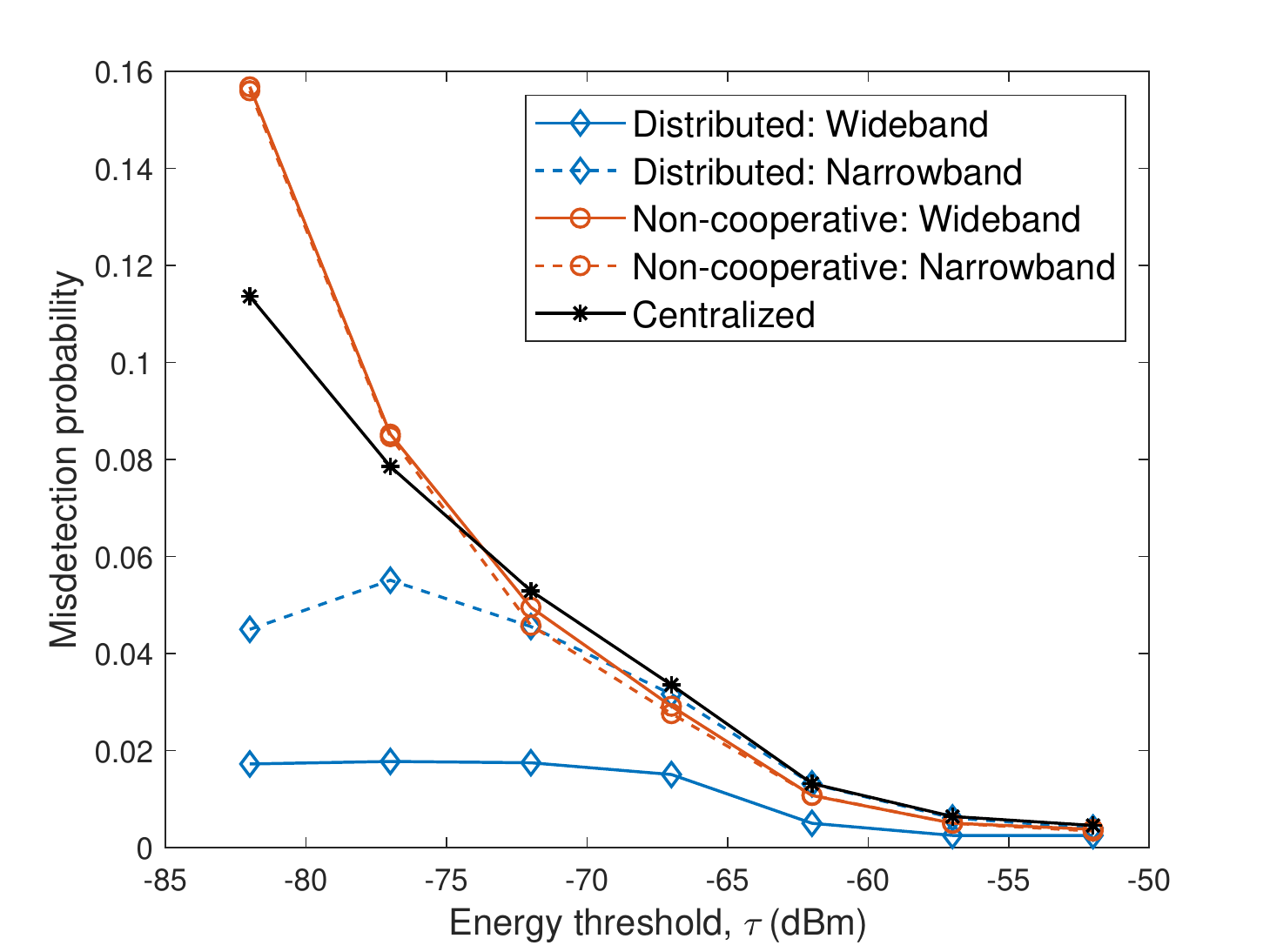}
		\caption{Misdetection probability}
		\label{fig:MD_vs_threshold}
	\end{subfigure}
	\caption{Performance with variations of the energy threshold.} 
	\label{fig:varyTh}
\end{figure}

We first study the performance under different detection thresholds. Fig. \ref{fig:UT_vs_threshold} shows the utilization ratio with variations of the energy thresholds. Clearly, increasing the energy threshold relaxes the coexistence requirement of the IoT network and incumbents, and thus more resources can be reused over space and frequency. Comparing the different schemes, we make the following observations. First, both the proposed and non-cooperative wideband solutions identify the highest number of available resources, as each BS senses the entire spectrum at its location. However, the proposed solution significantly outperforms the non-cooperative one in terms of misdetection, as shown in Fig. \ref{fig:MD_vs_threshold}, since the latter may incorrectly decide a busy channel to be available in the presence of a fading channel and/or shadowing. This is not the case with the proposed diffusion algorithm as cooperation helps enhance the reliability of decisions and reduce the misdetection probability. Second, the distributed narrowband solution significantly outperforms its non-cooperative counterpart although both schemes enforce each BS to sense one channel. This follows because by the end of the diffusion-based sensing procedure, each BS will have occupancy information across all channels, whereas in the non-cooperative one each BS will be limited to the availability of the sensed channel. Finally, the centralized solution performs well when the detection threshold is high. For lower detection threshold, if a single BS observes high energy, then it can be sufficient to bias all other BSs in the cluster to declare the channel to be busy. 

We also show snapshot of the spatial energy footprints across all channels, and the corresponding decisions at each BS for every scheme during that realization, where $\tau=-62$dBm. For the narrowband schemes, we highlight BSs with assigned channels in a different color. We also show the utilization ratio of the different schemes in that specific snapshot. It is evident that the distributed narrowband algorithm identifies the majority of available resources although each BS is assigned a single band. 

\begin{figure*}[t!]
	\centering
	\begin{subfigure}[t]{.32\textwidth}
		\centering
		\includegraphics[width=2.6in]{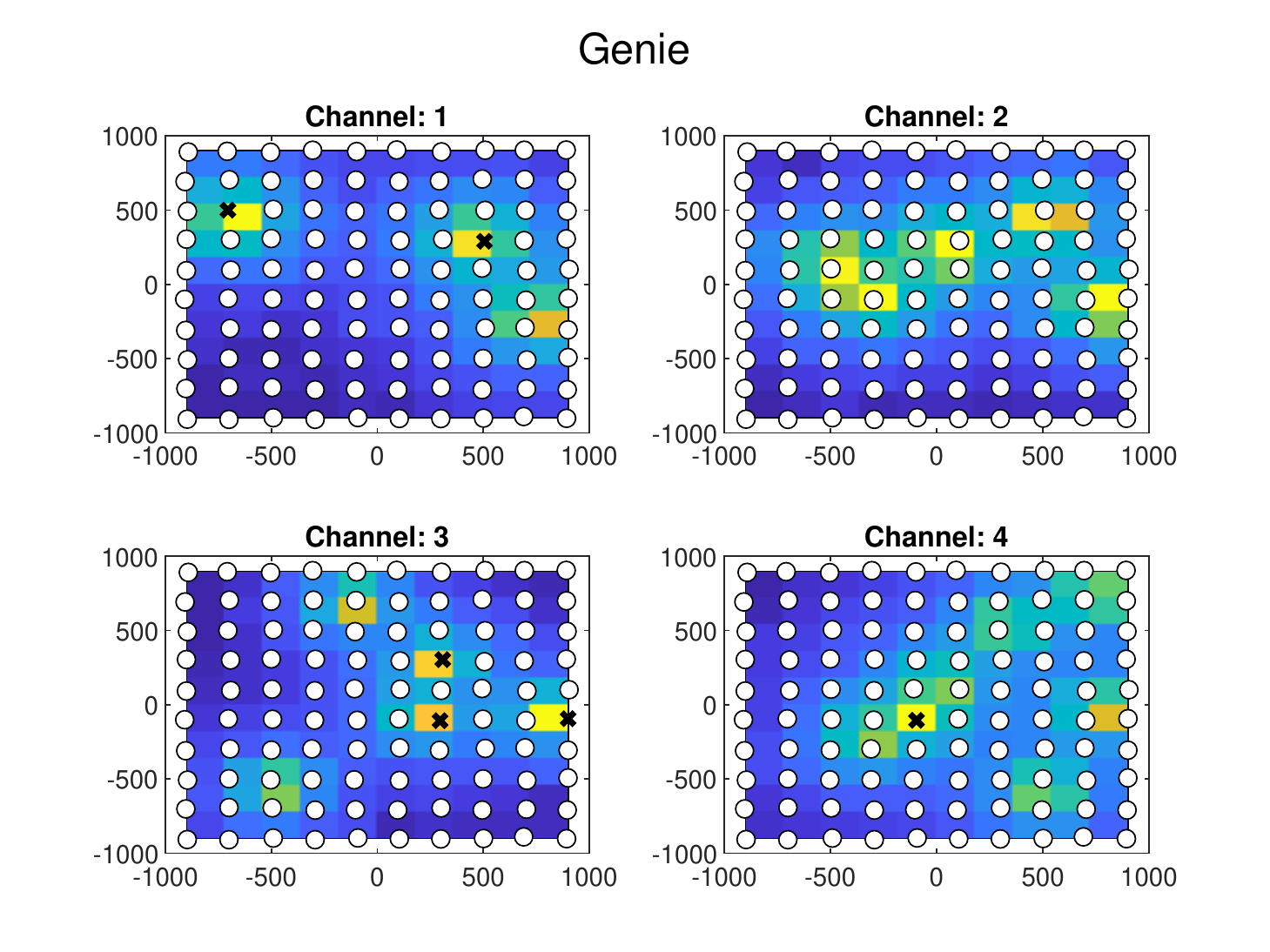}
		\caption{Genie system}
		\label{fig:footprint_Genie}
	\end{subfigure}~
	\begin{subfigure}[t]{.32\textwidth}
		\centering
		\includegraphics[width=2.6in]{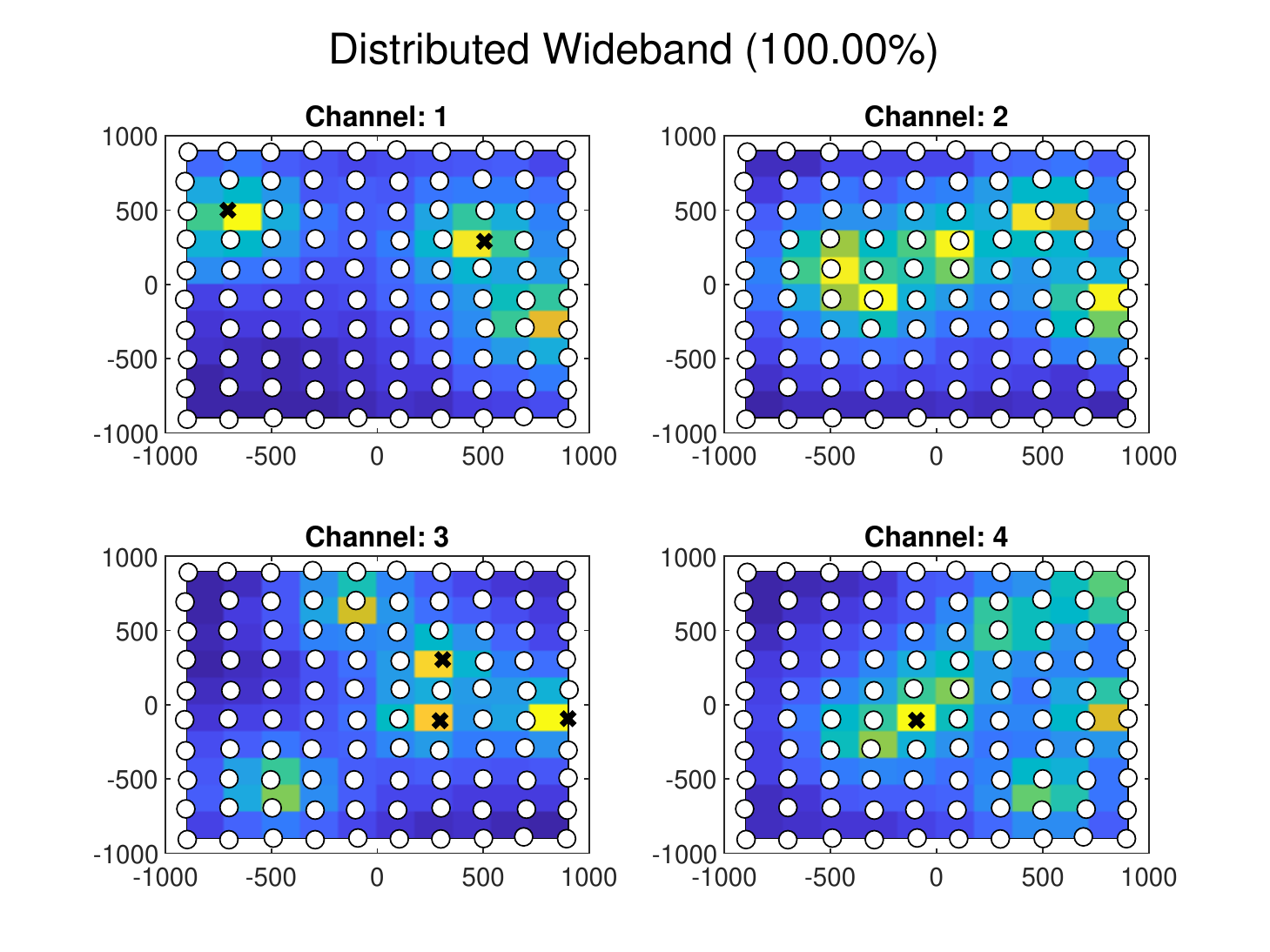}
		\caption{Distributed wideband}
		\label{fig:footprint_Prop_MB}
	\end{subfigure}~
	\begin{subfigure}[t]{.32\textwidth}
		\centering
		\includegraphics[width=2.6in]{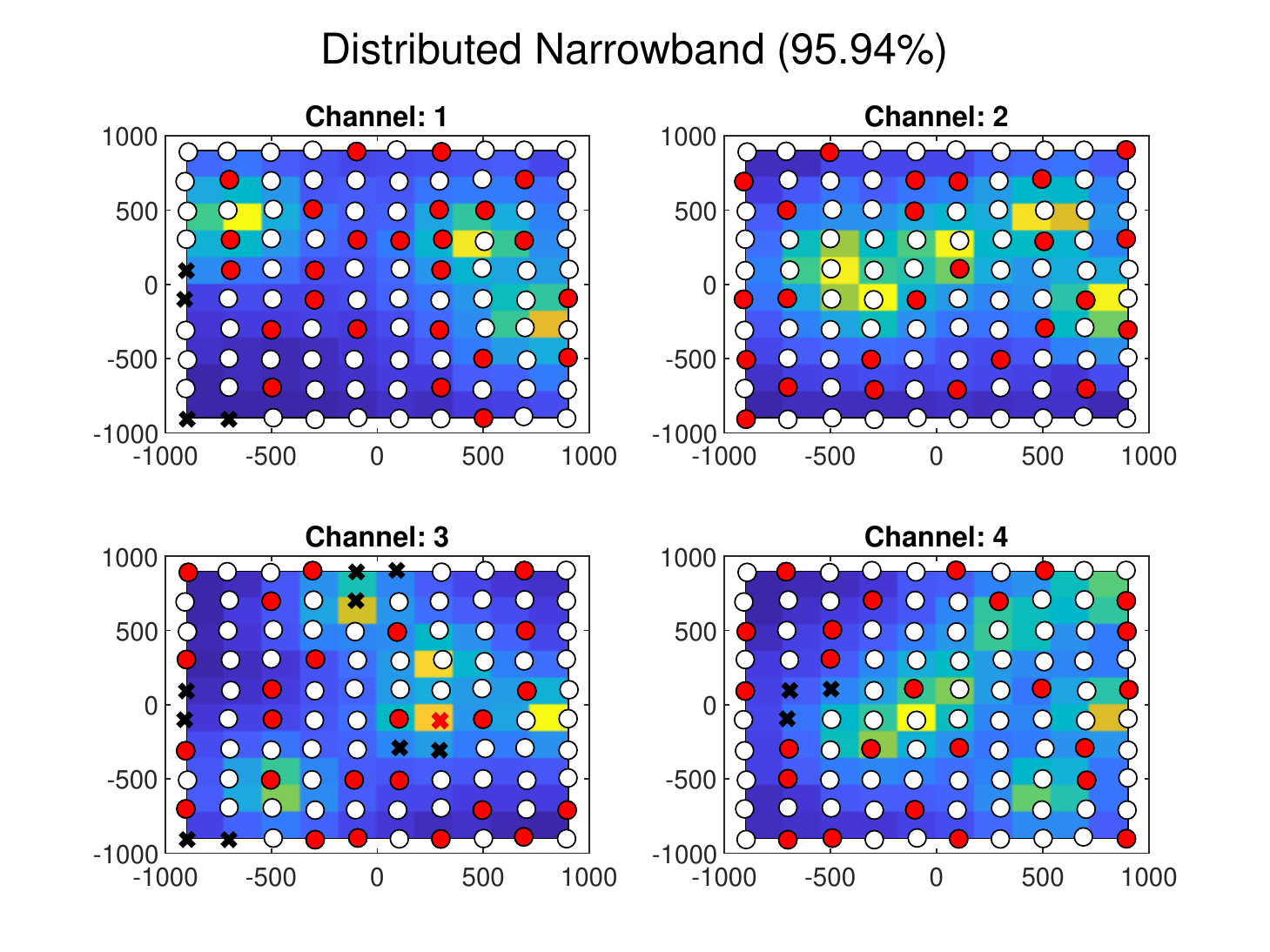}
		\caption{Distributed narrowband}
		\label{fig:footprint_Prop_SB}
	\end{subfigure}\\
	\begin{subfigure}[t]{.32\textwidth}
		\centering
		\includegraphics[width=2.6in]{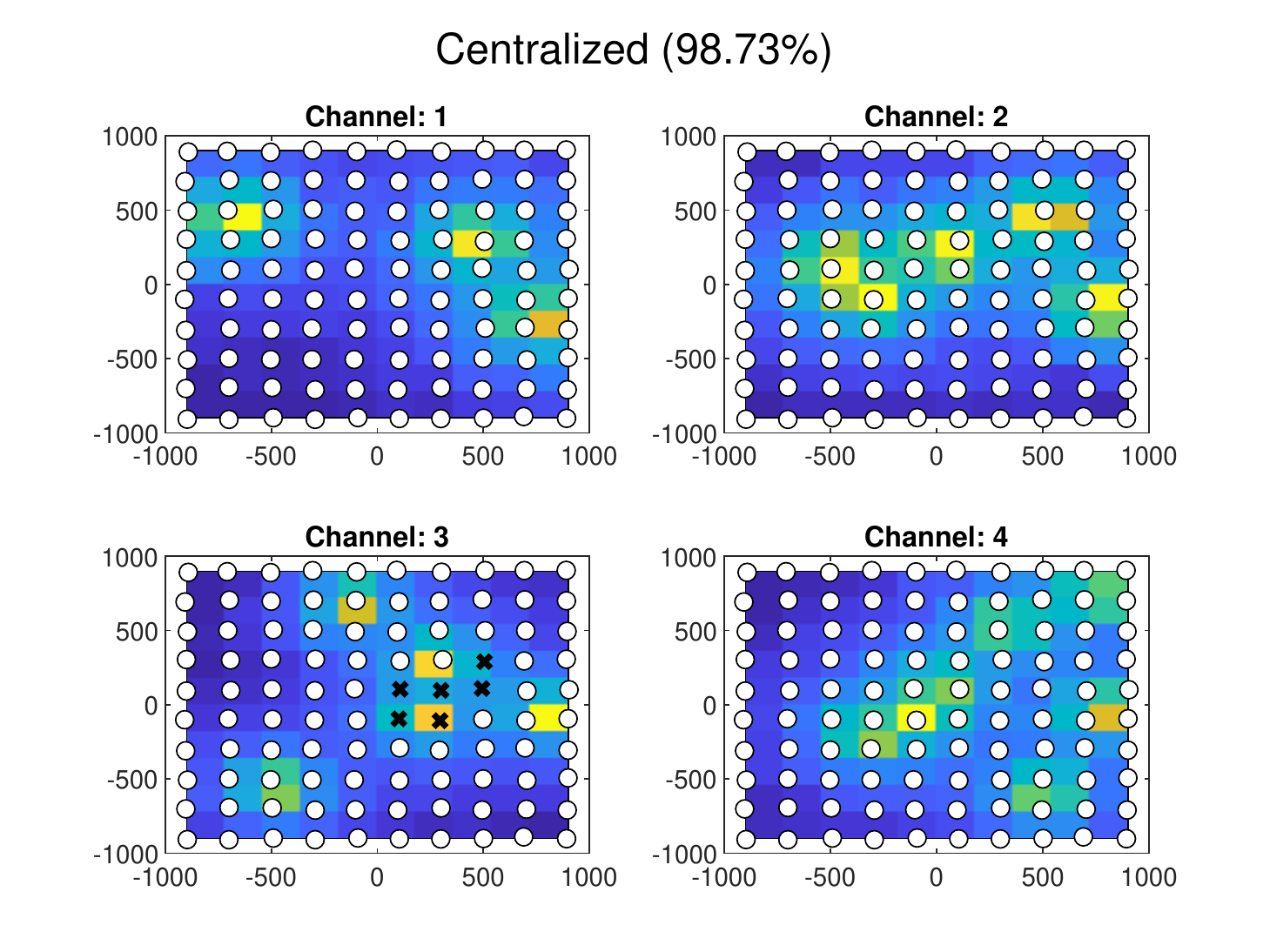}
		\caption{Centralized}
		\label{fig:footprint_Centralized}
	\end{subfigure}~
	\begin{subfigure}[t]{.32\textwidth}
		\centering
		\includegraphics[width=2.6in]{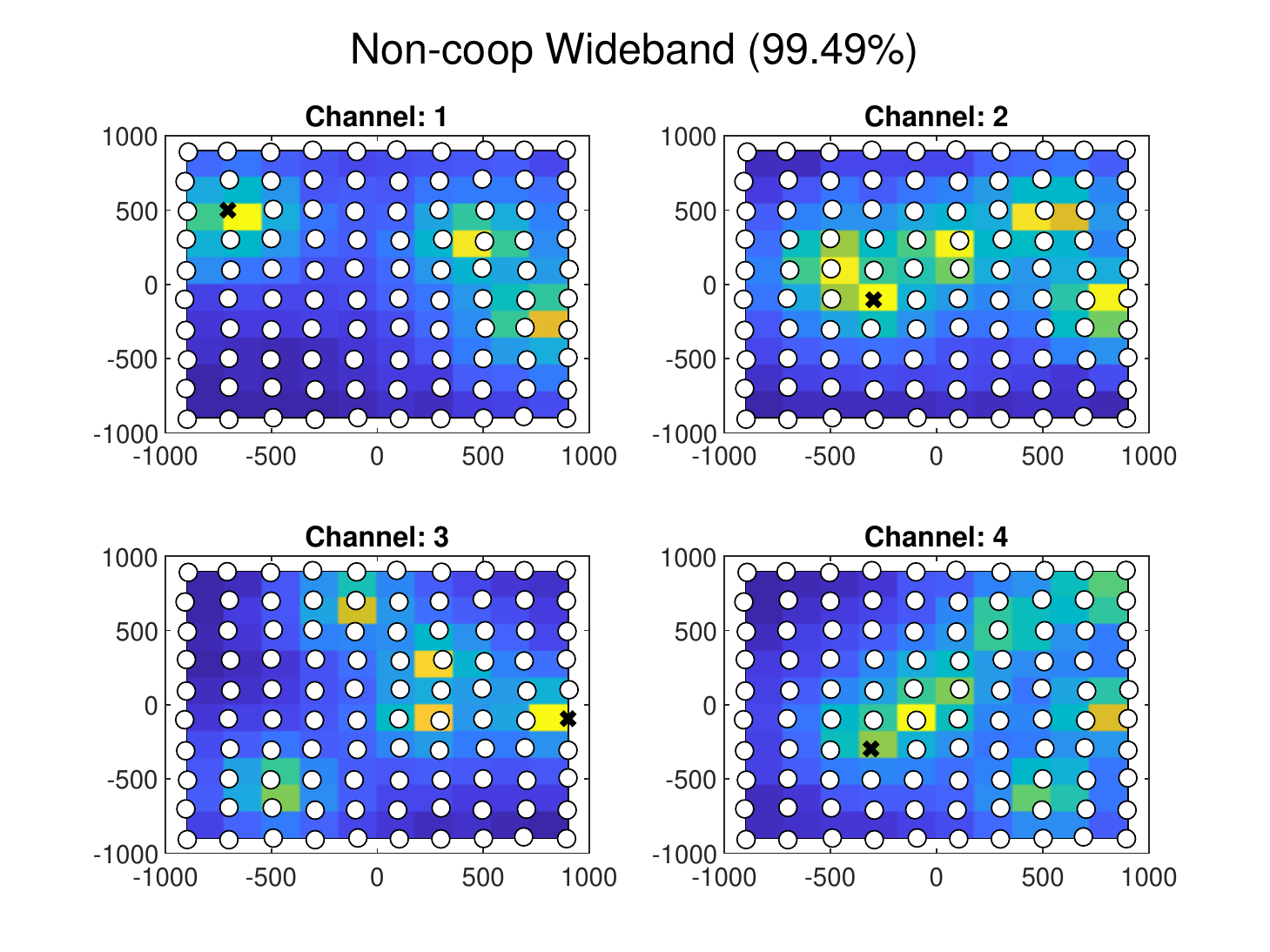}
		\caption{Non-cooperative wideband}
		\label{fig:footprint_NonCoop_MB}
	\end{subfigure}~
	\begin{subfigure}[t]{.32\textwidth}
		\centering
		\includegraphics[width=2.6in]{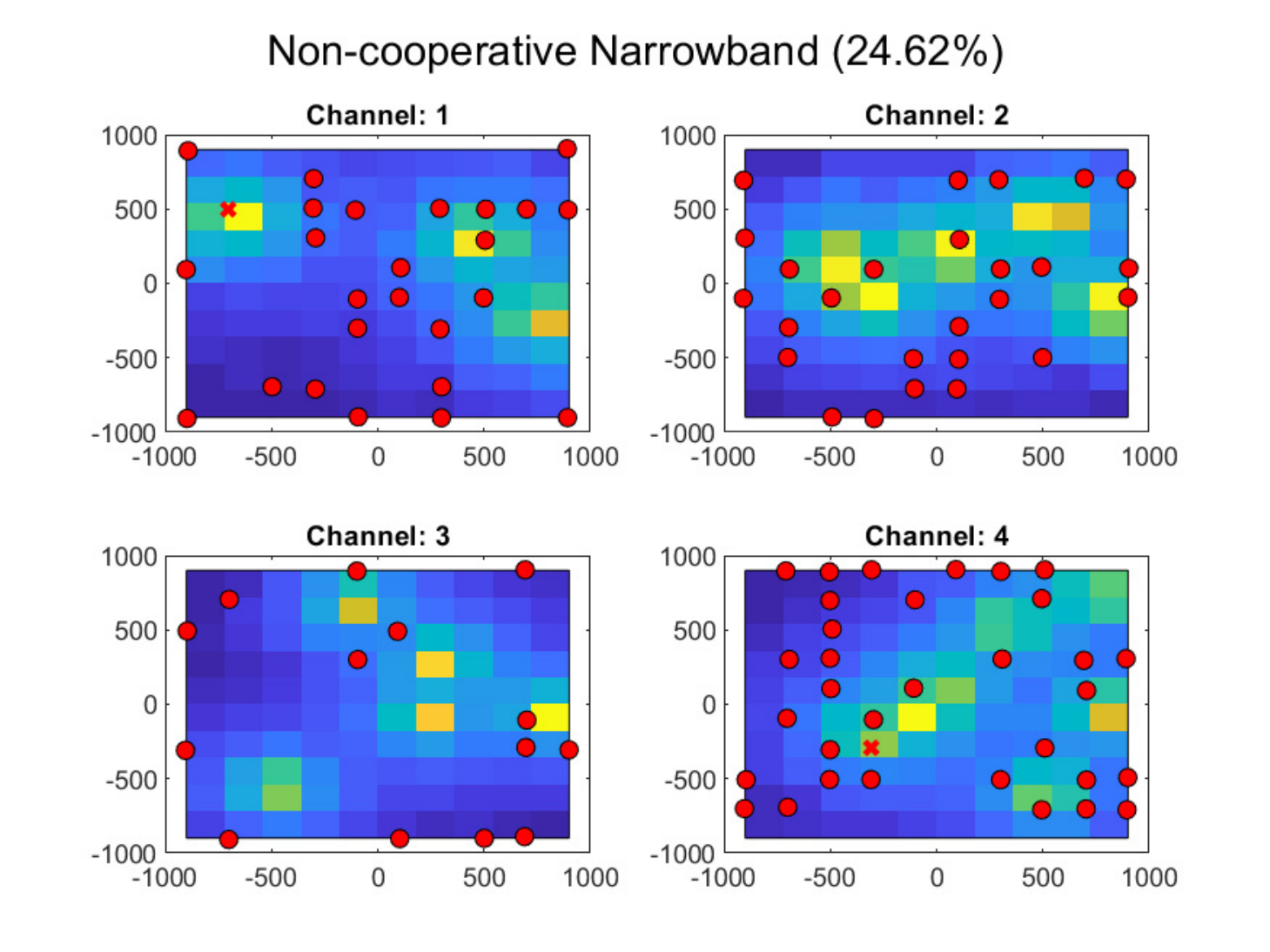}
		\caption{Non-cooperative narrowband}
		\label{fig:footprint_NonCoop_SB}
	\end{subfigure}
	\caption{The spatio-spectral footprints. `$\circ$' and `$\times$' denote available and busy decisions at their locations, respectively. For narrowband systems, BSs with their assignments are highlighted in a different color.} 
	\label{fig:footprints}
	\vspace{-.1in}
\end{figure*}

\subsubsection{Impact of communication radius}
\begin{figure}[t!]
	\centering
	\begin{subfigure}[t]{.4\textwidth}
		\centering
		\includegraphics[width=3in]{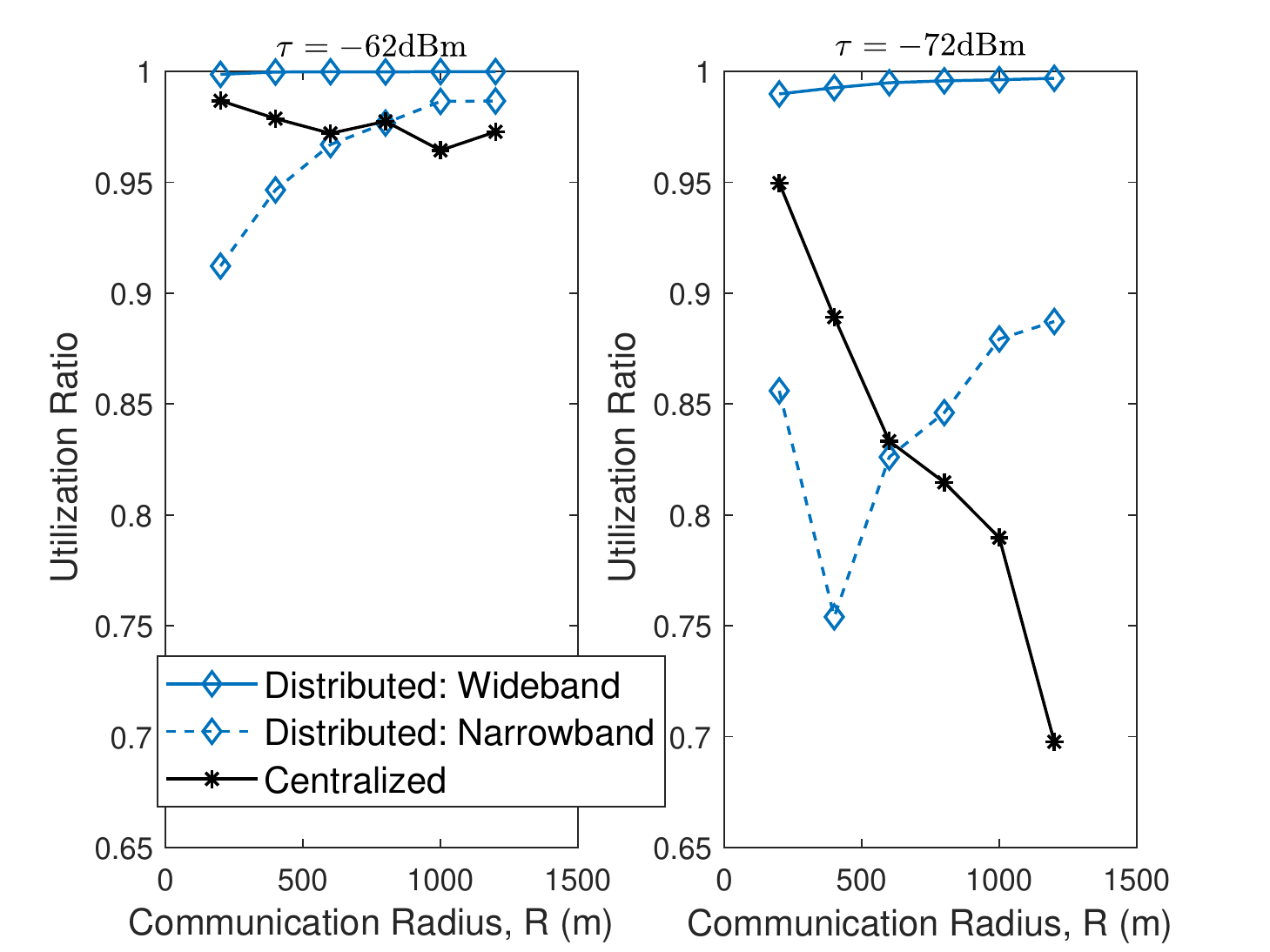}
		\caption{Utilization ratio}
		\label{fig:UT_vs_radius}
	\end{subfigure}~
	\begin{subfigure}[t]{.4\textwidth}
		\centering
		\includegraphics[width=3in]{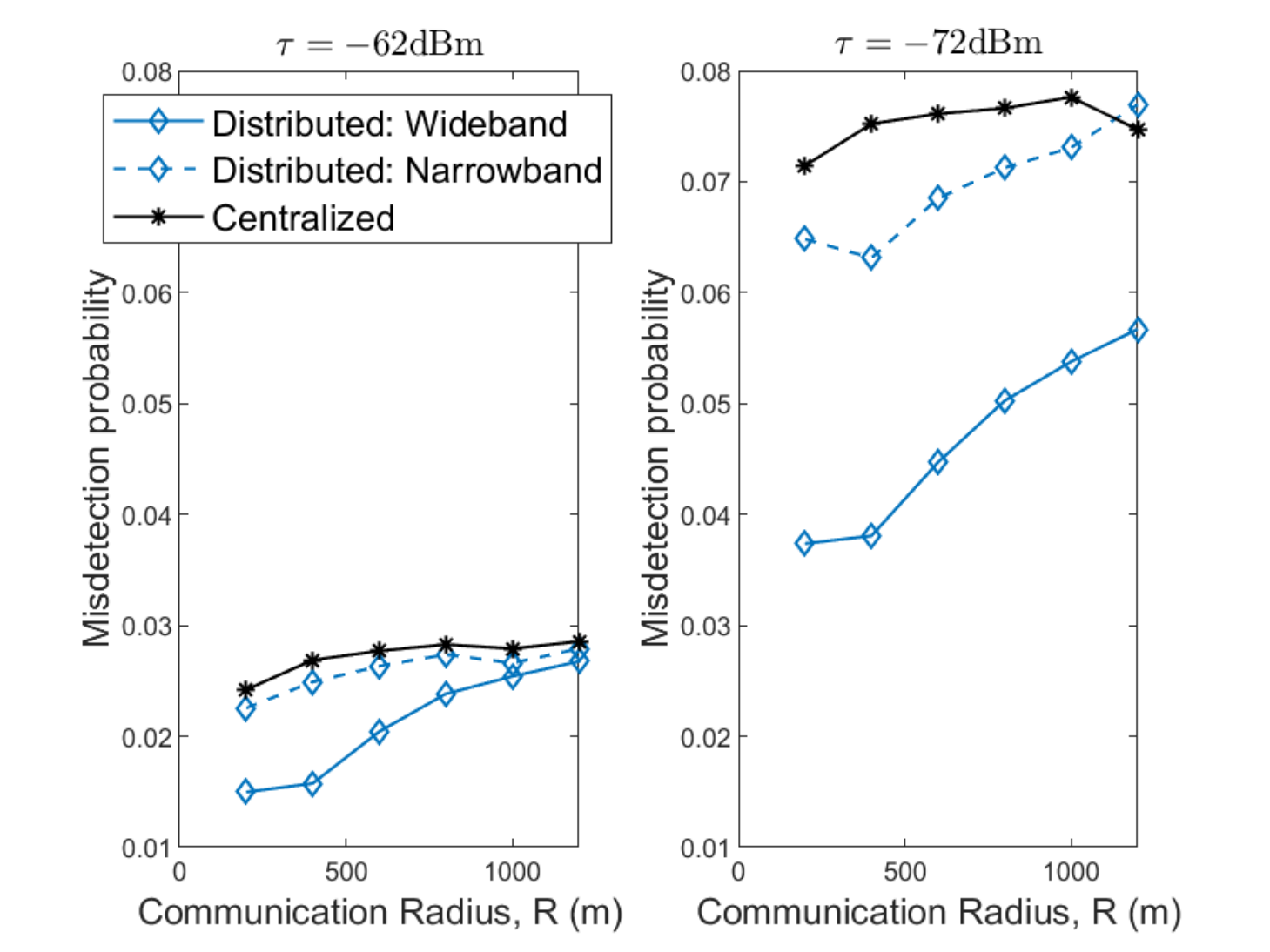}
		\caption{Misdetection probability}
		\label{fig:MD_vs_radius}
	\end{subfigure}
	\caption{Performance of cooperative schemes with variations of communication radius.} 
\end{figure}

We study the performance of the cooperative algorithms in terms of the communication radius $R$, which affects the neighborhood and cluster sizes  for the distributed and centralized sensing algorithms, respectively. We note that for a given $R$,  we find the average number of neighbors, and use it to determine how many clusters to use for the centralized scheme. In Fig. \ref{fig:UT_vs_radius}, we show the utilization ratio for different communication radii. It is observed that increasing $R$ for the distributed sensing algorithm does not destroy the spatial information, but in the contrary, it improves the utilization ratio, particularly for the narrowband system. This follows because each BS collects more sensing reports,  while still using adaptive weights throughout the sensing process to ensure that reports from nearby BSs are of higher value to those collected from far BSs. Such improvements come at the expense of slight increase in misdetection as shown in Fig. \ref{fig:MD_vs_radius}. This follows because a BS that receives a weak incumbent signal could decide the incumbent is inactive when many of its neighboring BSs do not detect the incumbent. In contrast, not only misdetection increases in the centralized scheme, but its utilization ratio also decreases with $R$. This follows because  all members of the cluster arrive to a single global decision, and thus as the cluster size increases, the spatial resolution decreases, with tangible reduction in the utilization ratio when the detection threshold is very low, e.g., $-72$dBm.

\subsection{Performance evaluation of the distributed resource allocation algorithm}

\subsubsection{Set-up}
We use the previous set-up, where now we randomly drop a very large number of IoT devices in the network so that each BS has at least four IoT devices to schedule.   We assume the transmit power of IoT devices to be $14$dBm. We then consider the following schemes:
\begin{itemize}
	\item \textbf{Centralized wideband and narrowband}: We use the sensing decisions of the distributed wideband and narrowband sensing algorithms, respectively, and then solve  (\ref{eq:centralizedAllocation}). 
	\item \textbf{Distributed wideband and narrowband}: We use Alg. \ref{alg:allocation} for resource allocation. The sensing decisions are obtained from the wideband and narrowband  sensing algorithms, respectively. 	
	\item \textbf{Non-cooperative narrowband}: Each BS picks the channel it sensed if it is deemed available. 
\end{itemize}  

\subsubsection{Impact of network densification}
\begin{figure}[t!]
	\centering
	\begin{subfigure}[t]{.32\textwidth}
		\centering
		\includegraphics[width=2.6in]{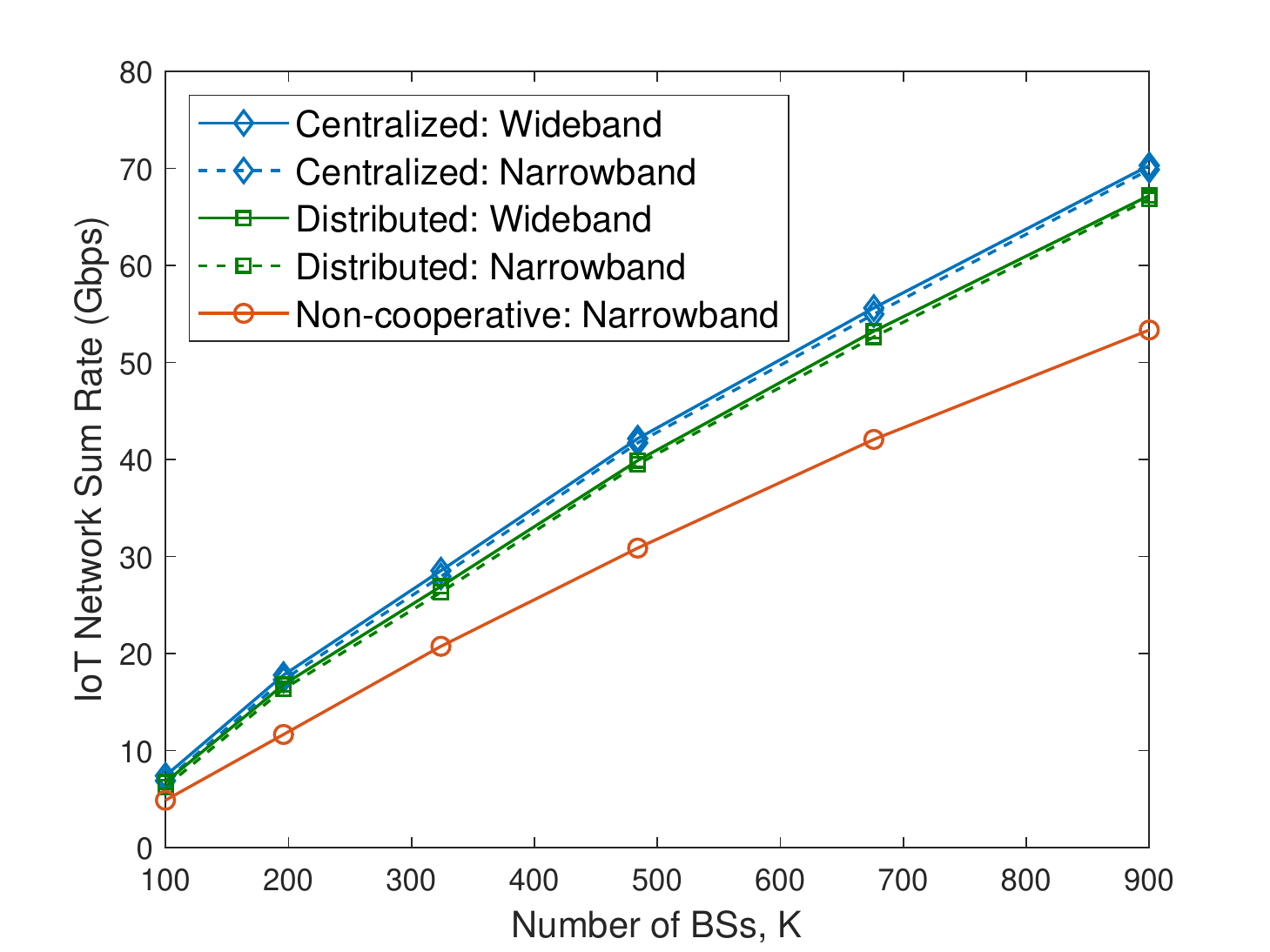}
		\caption{Sum rate of IoT network}
		\label{fig:sumRate_vs_numBss}
	\end{subfigure}~~
	\begin{subfigure}[t]{.32\textwidth}
		\centering
		\includegraphics[width=2.6in]{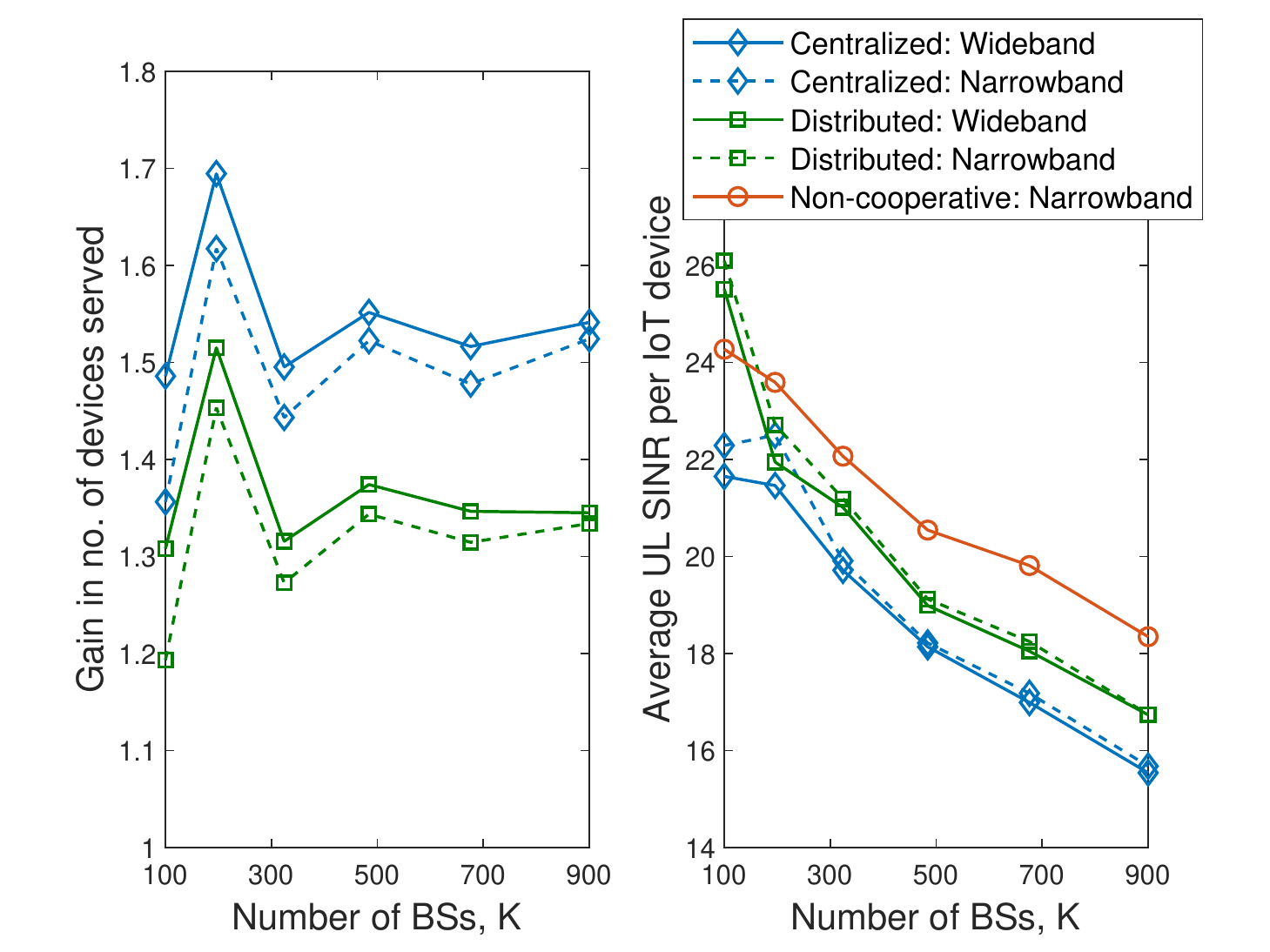}
		\caption{Gain in number of devices served and UL SINR per device }
		\label{fig:SINR_vs_numBSs}
	\end{subfigure}~~
	\begin{subfigure}[t]{.32\textwidth}
		\centering
		\includegraphics[width=2.6in]{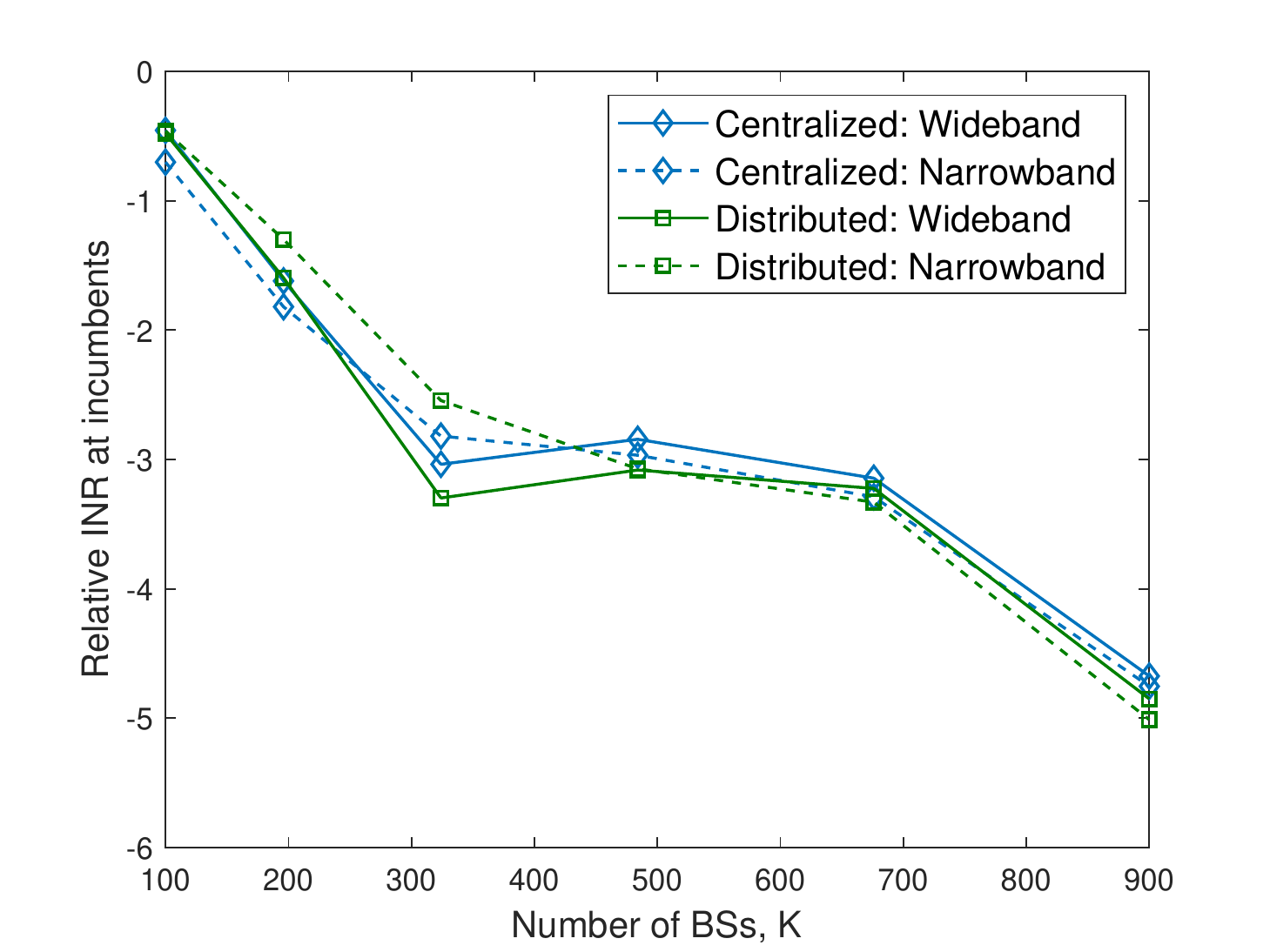}
		\caption{Relative INR at incumbents}
		\label{fig:INR_vs_numBSs}
	\end{subfigure}
	\caption{Comparison of the different spatio-spectral resource allocation schemes.}
	\label{fig:accessPerformance}
\end{figure}

We study performance of the different schemes when we increase the density of   BSs, which is accomplished by reducing the inter-site distance between them.  In Fig. \ref{fig:sumRate_vs_numBss}, we show the IoT network sum rate across the 80MHz spectrum. As expected, the centralized framework achieves the highest network capacity because it has global knowledge of the sensing decisions and rewards at each BS. The wideband implementation is slightly better than its narrowband counterpart as the former has very reliable sensing performance (cf. Fig. \ref{fig:varyTh}). More importantly, the distributed resource allocation performs relatively close to the centralized implementation. Overall, the proposed schemes significantly improves the IoT network sum rate compared to the non-cooperative allocation. This can be explained via Fig. \ref{fig:SINR_vs_numBSs}. In particular, we show the gain of the proposed schemes over the non-cooperative scheme in terms of the average number of IoT devices served in the network. Focusing on the schemes with narrowband sensing, it is shown that in dense networks, the centralized and distributed narrowband schemes increase the number of devices by 1.5x and 1.3x, respectively, compared to the non-cooperative scheme. Equally important, the mean SINR for the IoT device only slightly degrades under the proposed schemes, explaining the overall increase in the IoT network capacity. For example, when the density of BSs is 900, the distributed narrowband scheme adds approximately 300 devices to the network compared to the non-cooperative scheme, yet the SINR per IoT device under the former is only 1.6dB lower than that under the non-cooperative scheme. 

In Fig.  \ref{fig:INR_vs_numBSs}, we show the INR at the incumbents relative to that when non-cooperative allocation is used. It is evident that the incumbents are more protected under the proposed scheme compared to the non-cooperative one, particularly under high network densification. For example, the INR is reduced by approximately 5dB under the distributed narrowband scheme when the BS density is 900 BSs. To summarize, Fig. \ref{fig:accessPerformance} illustrates that both the IoT network and the incumbent network benefit from the proposed allocation schemes.

\subsection{Case study: IoT for public parks} 
\begin{figure}[t!]
	\centering
	\begin{subfigure}[t]{.4\textwidth}
		\centering
		\includegraphics[width=3in]{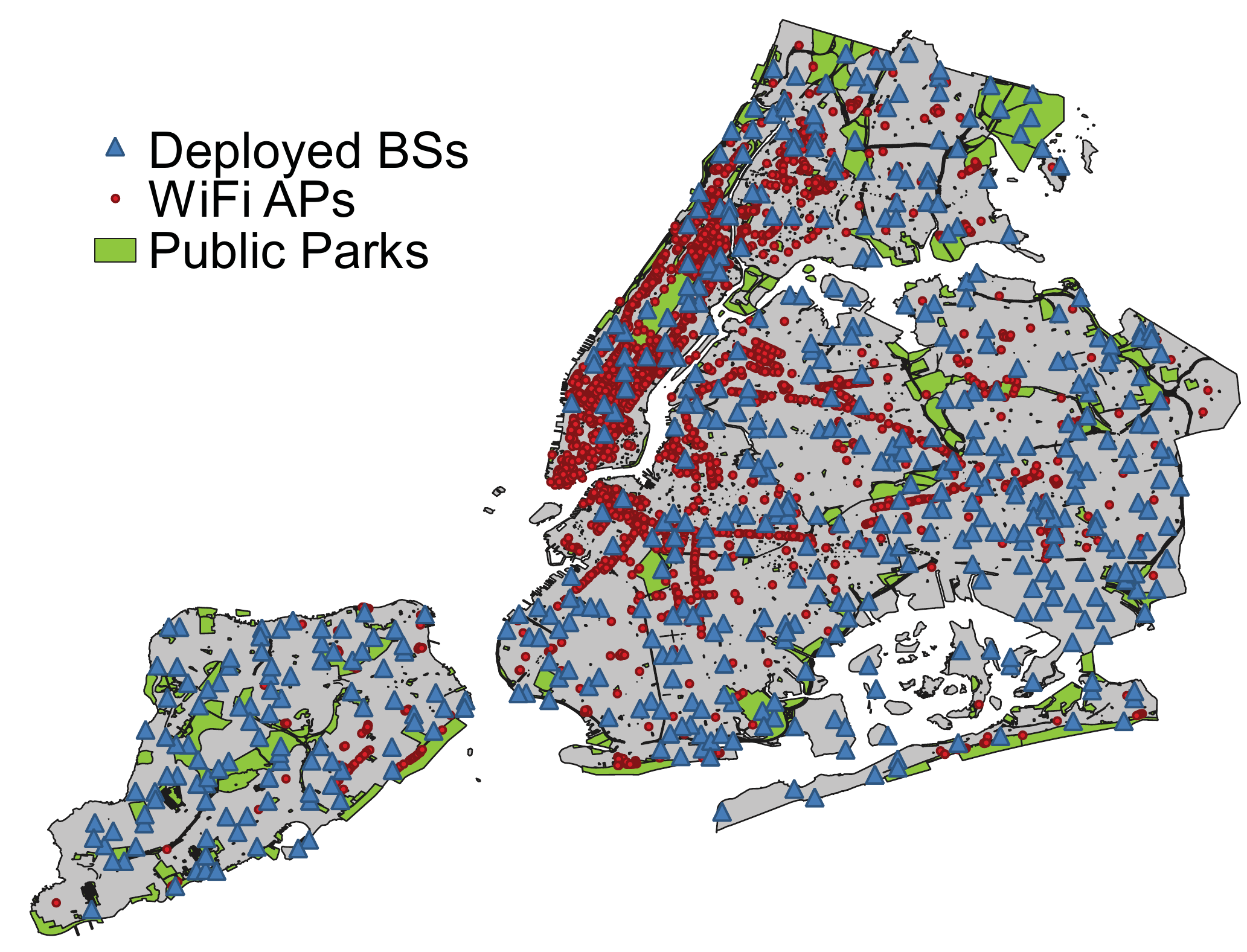}
		\caption{An illustration of the large-scale network set-up.}
		\label{fig:NYC}
	\end{subfigure}~
	\begin{subfigure}[t]{.4\textwidth}
		\centering
		\includegraphics[width=3in]{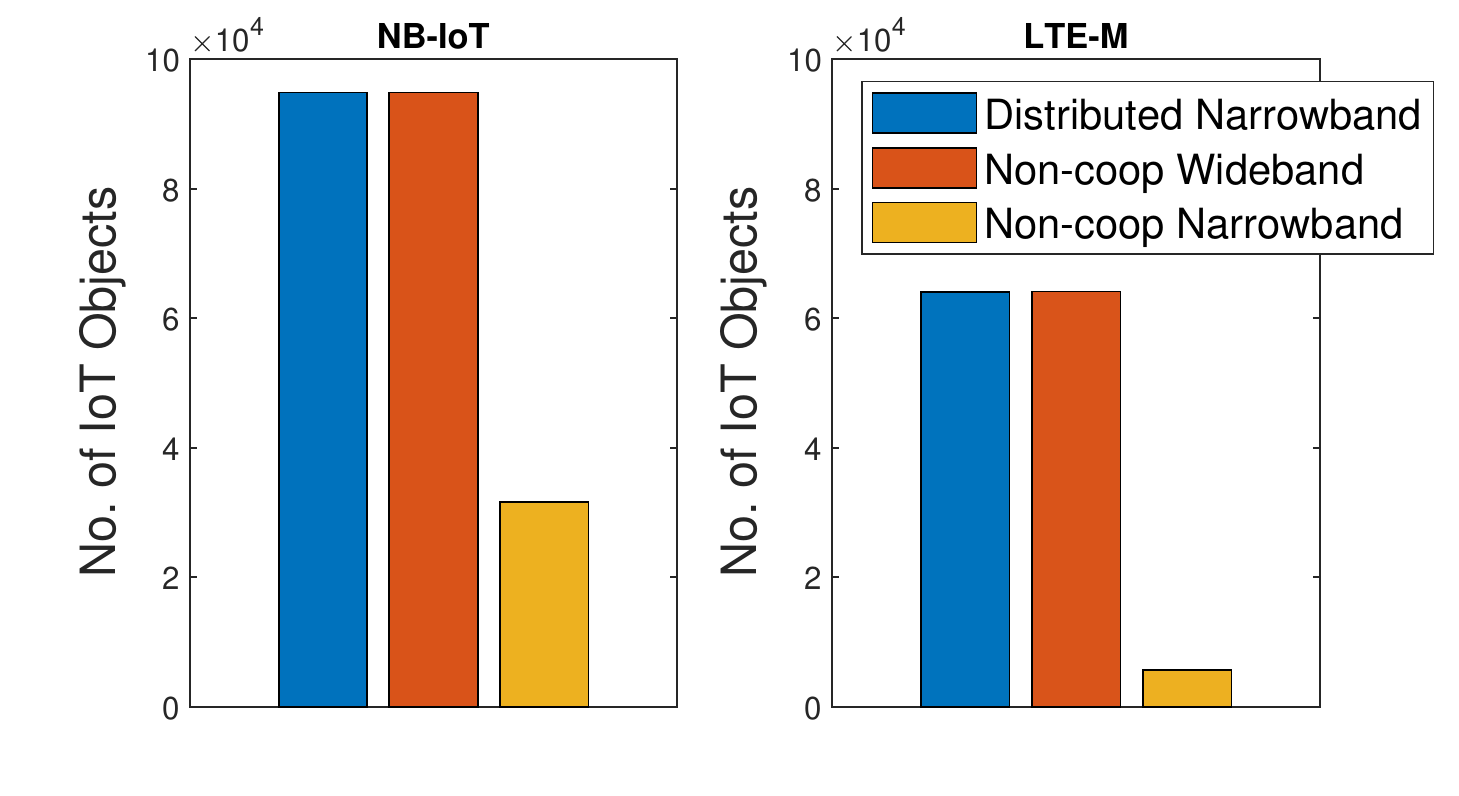}
		\caption{Average number of scheduled devices per scheme.}
		\label{fig:CaseStudySim}
	\end{subfigure}
	\caption{Case study for massive IoT in public parks.}
\end{figure}

We test the proposed distributed narrowband system in a more realistic setting, where we consider a massive IoT application for the public parks in New York City, where sensors and machines can be deployed for water management, tracking traffic activities in the parks, etc. We randomly drop 500 BSs across the city and use the \emph{NYC Open Data} to extract the locations of 2000 outdoor public WiFi APs, where we use their exact coordinates. We treat these APs as incumbents that use the same spectrum. We consider two IoT operations: LTE-M and NB-IoT, where the spectrum is channelized into 1.4MHz and 180KHz channels, respectively. Each WiFi AP is assumed to transmit a signal at 30dBm over the signal bandwidth, which is either 20MHz, 40MHz, or 80MHz within the wideband spectrum. In the proposed system, the assignment scheduler, i.e., Alg. \ref{alg:scheduler}, assigns each BS a subset of channels with a total bandwidth of 20MHz. Finally, we randomly drop $1\times 10^5$ IoT objects in the public parks, and assume all require access to the network. An illustration of the network is given in Fig. \ref{fig:NYC}.  

Fig. \ref{fig:CaseStudySim} shows the average number of devices that are scheduled over channels that are correctly identified as available for each scheme. It is observed that for NB-IoT, almost all devices are scheduled via the proposed and the non-cooperative wideband schemes. It is also clear that the proposed system significantly outperforms the non-cooperative narrowband system, emphasizing that cooperation across neighboring BSs is not only beneficial to enhance the reliability of sensing a particular channel, but also useful to infer the occupancy of other channels, given a proper sensing assignment. Similar trends hold for LTE-M, yet fewer devices are scheduled in comparison to NB-IoT, as each device requires larger bandwidth. 

\section{Conclusions}\label{sec:conclusion}
Providing massive IoT connectivity in the unlicensed spectrum requires exploring a large pool of narrowband channels and aggressively reusing them over space. To obtain a high spatio-spectral map of available resources, sensing a wideband spectrum at a fine resolution is required, while distributed processing of sensing reports helps capture the spatial variations of incumbents' footprints. In this paper, we propose a sensing assignment scheduler to limit the sensing burden on each BS. In particular, the proposed scheduler ensures that every BS is surrounded by other BSs sensing other parts of the spectrum. We further propose a distributed sensing algorithm, where BSs locally share and process sensing reports, improving the sensing reliability while maintaining the spatial resolution thanks to the online clustering running in parallel to sensing. When each BS obtains the spatio-spectral map at its location, distributed allocation of these resources across BSs is essential to limit intra- and inter-network interference. To this end, we have proposed a fast distributed allocation algorithm that ensures a channel is used by a single BS in a given neighborhood. Simulation results have demonstrated the effectiveness of the proposed architecture in comparison with centralized and non-cooperative schemes. It is shown that distributed sensing helps identify more spatio-spectral blocks at lower misdetection. As a consequence, distributed resource allocation enables connecting more IoT devices, while providing higher protection to incumbents. 

\appendix
We first show that $d_{k,m}$ has the same mean as $Y_{k,m}$, yet with lower variance. Indeed, we can rewrite (\ref{eq:d_ED}) recursively as $d_{k,m,i} = \zeta^i d_{k,m,0} + (1-\zeta) \sum_{l=0}^{i-1} \zeta^l Y_{k,m,i-l}$. Let $d_{k,m,0}=Y_{k,m,0}$, and assume that $\{Y_{k,m,l}\}_{l=0}^i$ are independent and identically distributed random variables, then we have
\begin{equation}
\begin{aligned}
\mathbb{E}[d_{k,m,i}] &= \zeta^i \mathbb{E}[Y_{k,m,0}] + (1-\zeta) \mathbb{E}[Y_{k,m,i-1}]\sum_{l=0}^{i-1} \zeta^l ~\stackrel{(i\rightarrow\infty)}{=}  \mathbb{E}[Y_{k,m}],
\end{aligned}
\end{equation}
where in the last equality we consider the steady-state behavior. Similarly, the variance in steady-state can be computed as
\begin{equation}
\begin{aligned}
\mathbb{V}(d_{k,m,i}) &\stackrel{(i\rightarrow\infty)}{=}  (1-\zeta)^2 \mathbb{V}\left(\sum_{l=0}^{i-1} \zeta^l Y_{k,m,i-l}\right)~
= \frac{(1-\zeta)^2}{1-\zeta^2}  \mathbb{V}(Y_{k,m}), 
\end{aligned}
\end{equation}
which is much smaller than $\mathbb{V}(Y_{k,m})$. 

Next, we compute $\mathbb{E}_{0} [\boldsymbol w_i]$ when $Y_{k,i}$ is the energy of a Gaussian noise with variance $P_n$. To this end, it can be shown  $\mathbb{E}_{0}[\mathbf{Y}_i]=3P_n^2\mathbf{I}$. Similarly, the $k$-th entry in $\mathbb{E}_{0}[\boldsymbol{\phi}_i]$ is $\mathbb{E}_0[Y_{k,i}d_{k,i}]=(3-2\zeta)P_n^2$. To compute $\mathbb{E}_1 [\boldsymbol w_i]$ when $Y_{k,i}$ is the energy of a Gaussian sample with variance $P_n$ and mean $\sqrt{P_s}$, we have $\mathbb{E}_{1}[\mathbf{Y}_i]=(P_s^2+6P_sP_n+3P_n^2)\mathbf{I}$ and $\mathbb{E}_1[Y_{k,i}d_{k,i}]=(1-\zeta)(P_s^2+6P_sP_n+3P_n^2)+\zeta(P_s+P_n)^2$.

To compute the variance, we can expand $\tilde{\boldsymbol{w}}_i^T\boldsymbol{\Sigma}\tilde{\boldsymbol{w}}_i$, and then take the expectation on both sides to get (\ref{eq:EnergyConservation}). Note that in this case, $\mathbf{B}= (\mathbf{I}-3P_n^2\mathbf{M})\mathbf{A}^T$. Also, expanding the right-hand side of (\ref{eq:EnergyConservation}). we get
\begin{equation}
\begin{aligned}
\mathbb{E}_0[(\boldsymbol{\phi}_i-\mathbf{Y}_i \bar{\boldsymbol{w}_i})^T\mathbf{M}\boldsymbol{\Sigma}\mathbf{M}(\boldsymbol{\phi}_i-\mathbf{Y}_i \bar{\boldsymbol{w}_i})] =
\operatorname{Tr}(\mathbf{M}\boldsymbol{\Sigma}\mathbf{M} \bar{\mathbf{Z}} ) + \bar{\boldsymbol{w}_i}^T \mathbf{M} \mathbb{E}_0[\mathbf{Y}_i\mathbf{\Sigma} \mathbf{Y}_i] \mathbf{M}\bar{\boldsymbol{w}_i} - 2 \bar{\boldsymbol{w}_i}^T \bar{\mathbf{c}},
\end{aligned}
\end{equation}
where $\operatorname{Tr}(\cdot)$ is the trace operation, $\bar{\mathbf{Z}}=\mathbb{E}_0[\boldsymbol{ \phi}_i\boldsymbol{ \phi}_i^T]$, and $\bar{\mathbf{c}}=\mathbb{E}_0[\mathbf{Y}_i\mathbf{M\Sigma M\boldsymbol{ \phi}}_i]$. It can be shown that the $k$-th diagonal element of  $\bar{\mathbf{Z}}$ is $\mathbb{E}_0[ \phi_{k,i}^2]=(1-\zeta)^2 P_n^4 (105+\frac{30}{1-\zeta}+\frac{9\zeta^2}{1-\zeta^2}+\frac{6\zeta^3}{(1+\zeta)(1-\zeta^2)})$, and the off-diagonal $(l,j)$-th element is $\mathbb{E}_0[\phi_{l,i}\phi_{j,i}]=(3-2\zeta)^2 P_n^4$.  Similarly, the $k$-th diagonal element of $\mathbb{E}_0[\mathbf{Y}_i\mathbf{\Sigma} \mathbf{Y}_i] $ is $\mathbb{E}_0[Y_{k,i}^4] \sigma_{k,k} = 105 P_n^4\sigma_{k,k}$, where $\sigma_{k,k}$ is the $k$-th diagonal element of $\mathbf{\Sigma}$. Also, the off-diagonal $(l,j)$-th element is $\mathbb{E}_0[Y_{l,i}^2 Y_{j,i}^2] \sigma_{l,j}= 9 P_n^4 \sigma_{l,j}$. Finally, to compute $\bar{\mathbf{c}}$, let $\boldsymbol{\theta}_k=[\mu_1\mu_k \sigma_{1,k},\mu_2\mu_k \sigma_{2,k},\cdots,\mu_K\mu_k \sigma_{K,k}]$, i.e., the $k$-th column vector of $\mathbf{M\Sigma M}$. Then, the $k$-th element of  $\bar{\mathbf{c}}$ is 
\begin{equation}
\begin{aligned}
\mathbb{E}_0[\boldsymbol{ \phi}_i^T \boldsymbol{\theta}_k Y_{k,i}^2]&= \mathbb{E}_0\left[Y_{k,i}^2 \sum_{j=1}^K Y_{j,i} d_{j,i} \mu_j \mu_k\sigma_{j,k}\right]\\
&= \mu_k^2 \sigma_{k,k}\mathbb{E}_0[Y_{k,i}^3 d_{k,i}] + \mu_k\mathbb{E}_0[Y_{k,i}^2]  \mathbb{E}_0[Y_{j,i} d_{j,i}]\sum_{j=1,j\neq k}^K \mu_j \sigma_{j,k}\\
&= (105-90\zeta) \mu_k^2 \sigma_{k,k} P_n^4 + 3(3-2\zeta) P_n^4  \mu_k\sum_{j=1,j\neq k}^K \mu_j \sigma_{j,k}.
\end{aligned}
\end{equation}
Computing $\mathbf{\Sigma}$ from the discrete-time Lyapunov equation  $\boldsymbol{\Sigma}-\mathbf{B}\boldsymbol{\Sigma} \mathbf{B}^T=\mathbf{I}$, and plugging the aforementioned expressions in (\ref{eq:deflectionCoefficient}), we get the final theoretical expression for the deflection coefficient, which is shown to be in good agreement with Monte Carlo simulations in Fig. \ref{fig:theorticalDeflection}. 

\section*{Acknowledgment}
The authors would like to thank Zakarya El-friakh and Ljiljana Simić from the Institute for Networked Systems, RWTH Aachen University, for providing the RSS measurements for the indoor REM. 

\bibliographystyle{IEEEtran}
\bibliography{C:/Users/ghait/Dropbox/References/IEEEabrv,C:/Users/ghait/Dropbox/References/References}

\end{document}